\documentclass[%
 reprint,
 showpacs,
 nofootinbib,
 amsmath,amssymb,
 aps,
]{revtex4-1}

\newcommand{\initiallabel}{\mathrm{i}} 
\newcommand{\plancklabel}{\mathrm{p}}
\newcommand{\matterlabel}{\mathrm{m}}
\newcommand{\radiationlabel}{\mathrm{r}}
\newcommand{\eqlabel}{\mathrm{eq}}

\newcommand{\ti}{t_\initiallabel}
\newcommand{\phii}{\phi_\initiallabel}
\newcommand{\phidi}{\dot\phi_\initiallabel}
\newcommand{\Hi}{H_\initiallabel}

\newcommand{\phip}{\phi_\plancklabel}
\newcommand{\etap}{\eta_\plancklabel}
\newcommand{\Rp}{a_\plancklabel}
\newcommand{\phiz}{\phi_0}

\newcommand{\m}{m_\plancklabel}    

\newcommand{\tp}{t_\plancklabel}            
\newcommand{\tL}{t_\mathrm{\Lambda}}        
\newcommand{\etak}{\eta_\mathrm{\kappa}}    
\newcommand{\tm}{t_\matterlabel}            
\newcommand{\etar}{\eta_\radiationlabel}    

\newcommand{\rhomp}{\rho_\plancklabel^\matterlabel}
\newcommand{\rhorp}{\rho_\plancklabel^\radiationlabel}

\newcommand{\teq}{t_\eqlabel}               
\newcommand{\phieq}{\phi_\eqlabel}          
\newcommand{\phideq}{\dot{\phi}_\eqlabel}   

\newcommand{\HedS}{H_\mathrm{edS}}
\newcommand{\Hh}{H_\mathrm{h}}
\newcommand{\Hl}{H_\mathrm{l}}

\newcommand{\fm}{f_\mathrm{max}}
\newcommand{\fmn}{f_\mathrm{min}}
\newcommand{\fz}{f_0}
\newcommand{\hz}{h_0}
\newcommand{\psiz}{\psi_0}

\newcommand{\vellim}{\xi}

\newcommand{\Kfrac}{\mathcal{K}}

\newcommand{\Vpol}{V^\mathrm{pol}_n}         
\newcommand{\Vpow}{V^\mathrm{pow}_\epsilon}  
\newcommand{\Vm}{V_\mathrm{max}}             

\newcommand{\rhoph}{\rho_\phi}
\newcommand{\Pph}{P_\phi}

\newcommand{\Nflat}{N_\mathrm{e}}
\newcommand{\Nflatp}{\prm{N_\mathrm{e}}}
\newcommand{\Nflatpp}{\dprm{N_\mathrm{e}}}
\newcommand{\Hflat}{\dot{N}_\mathrm{e}}
\newcommand{\dHflat}{\ddot{N}_\mathrm{e}}

\def\tfrac#1#2{{\textstyle\frac{#1}{#2}}}

\newcommand{\difrac}[2]{\frac{d #1}{d #2}}    
\newcommand{\prm}[1]{{{#1}^\prime}}           
\newcommand{\dprm}[1]{{{#1}^{\prime\prime}}}  

\newcommand{\abs}[1]{\left|#1\right|}

\newcommand{\Sref}[1]{Section~\ref{#1}}        
\newcommand{\Fref}[1]{Figure~\ref{#1}}         

\renewcommand{\d}[2][]{\operatorname{d}^{#1}\!{#2}}

\newcommand{\smax}{\mathrm{max}}         
\newcommand{\bigO}[1]{{\sim\mathcal{O}{\left(#1\right)}}}
\newcommand{\Pnorm}{\mathcal{P}}
\newcommand{\Punnorm}{P}
\newcommand{\PR}{\Pnorm_{\mathcal{R}}}

\DeclareMathOperator{\arccosh}{arccosh}
\DeclareMathOperator{\arcsinh}{arcsinh}
\DeclareMathOperator{\arctanh}{arctanh}

\usepackage{graphicx}  

\begin{document}

\title{Kinetic initial conditions for inflation}
\author{W J Handley} 
\email{wh260@mrao.cam.ac.uk}

\author{S D Brechet} 
\email{sylvain.brechet@epfl.ch}

\author{A N Lasenby} 
\email{a.n.lasenby@mrao.cam.ac.uk}

\author{M P Hobson}
\email{mph@mrao.cam.ac.uk}

\affiliation{Astrophysics Group,
Cavendish Laboratory, J.~J.~Thomson Avenue, Cambridge, CB3 0HE, UK}


\date{\today}

%
\begin{abstract}
  We consider the classical evolution of the inflaton field $\phi(t)$
  and the Hubble parameter $H(t)$ in homogeneous and isotropic
  single-field inflation models.  Under an extremely broad assumption,
  we show that the Universe generically emerges from an initial
  singularity in a non-inflating state where the kinetic energy of the
  inflaton dominates its potential energy, $\dot{\phi}^2 \gg V(\phi)$.
  In this kinetically-dominated regime, the dynamical equations admit
  simple analytic solutions for $\phi(t)$ and $H(t)$, which are
  independent of the form of $V(\phi)$.  In such models, these
  analytic solutions thus provide a simple way of setting the initial
  conditions from which to start the (usually numerical) integration
  of the coupled equations of motion for $\phi(t)$ and $H(t)$.  We
  illustrate this procedure by applying it to spatially-flat models
  with polynomial and exponential potentials, and determine the
  background evolution in each case; generically $H(t)$ and
  $|\phi(t)|$ as well as their time derivatives decrease during
  kinetic dominance until $\dot{\phi}^2\sim V(\phi)$, marking the
  onset of a brief period of fast-roll inflation prior to a slow roll
  phase.  We also calculate the approximate spectrum of scalar
  perturbations produced in each model and show that it exhibits a
  generic damping of power on large scales.  This may be relevant to
  the apparent low-$\ell$ falloff in the CMB power spectrum.
\end{abstract}

\maketitle


\section{Introduction}

Cosmological inflation was first introduced by 
\citet{starobinskii_spectrum_1979}, \citet{guth_inflationary_1981} and 
others, and extended by \citet{linde_1982} and several other workers
to create modern inflationary theory. It is able to solve
long-standing problems with the paradigm of Big Bang cosmology. In
addition to solving the monopole, flatness and horizon problems,
inflation provides a mechanism for generating super-horizon scale
cosmological perturbations from quantum fluctuations of the inflaton
field (see, for example, \citet{mukhanov_theory_1992}). Inflation thus
predicts that large-scale structures in the Universe are the result of
quantum-mechanical fluctuations occurring during the inflationary
epoch. Inflationary perturbations of this type are consistent with the
anisotropy power spectrum of the cosmic microwave background (CMB)
\citep{hinshaw_nine-year_2012,planck_collaboration_planck_2013}.

In this paper, we focus primarily on the background dynamics of
single-field inflationary models, as determined by the evolution of
the scalar field \(\phi(t)\) and the Hubble parameter \(H(t)\) as
functions of cosmic time.  This cosmological evolution can generally
only be determined numerically, which requires initial conditions for
the numerical integration.  We therefore consider the limiting forms
of the coupled dynamical equations for \(\phi(t)\) and \(H(t)\) as one
evolves backwards in time and the universal scale factor \(a\to 0\). We 
work under the extremely broad assumption that there exists a time
prior to which \(|\dot{\phi}| > \vellim > 0\), for some positive
constant \(\vellim\), as \(a \to 0\).  With this assumption, we show
that as \(a\to 0\) the kinetic energy of the inflaton comes to dominate the 
potential energy: \(\dot{\phi}^2\gg V(\phi)\). We call this condition 
{\em kinetic dominance\/} (KD). This is generically true, except perhaps 
for a single special solution for each potential \(V(\phi)\).

Kinetically-dominated universes emerge from a singularity at a finite
time in the past and in a non-inflating state. This statement is true
even if additional auxiliary fluids are present such as radiation,
matter or curvature.  In the kinetically dominated regime, the coupled
equations of motion admit simple analytical solutions for \(\phi\) and
\(H\), which do not depend on the form of the inflaton potential
\(V(\phi)\).  These solutions therefore provide a simple way of setting 
the initial conditions for such inflation models.

With these initial conditions in hand, we then analyse (numerically)
the evolution of \(\phi(t)\) and \(H(t)\) in the flat case through to the 
end of inflation, thereby determining the background evolution, and
also calculate the spectrum of scalar perturbations produced. We find
that the latter generically has a cut-off at large spatial scales,
which could provide an explanation for the recently observed
low-\(\ell\) falloff in the CMB power spectrum
\citep{hinshaw_nine-year_2012,planck_collaboration_planck_2013}.
 
Throughout this paper we work many Planck times away from the
singularity: \(t\gg \tp\) so we do not expect quantum gravitational
effects to be present.  In addition, the homogeneity and scale of the
inflaton field indicates that the number of inflaton particles
\(n\gg1\), so quantum field theoretic effects will not be present.
Thus, although inflationary dynamics requires a rigorous quantum
treatment, it is possible to adopt a classical phenomenological
approach to setting initial conditions for the background dynamical
variables.  

The structure of this paper is as follows.  In
\Sref{sec:Scalar_field_inflation_models}, we will briefly introduce 
the dynamics of inflationary models based on a scalar field with the 
possibility of additional `auxiliary' fluids.  In
\Sref{sec:The_generic_nature_of_kinetic_dominance} we prove the
generic nature of kinetic dominance.  We explore the consequences of
kinetic dominance in \Sref{sec:consequences_of_kinetic_dominance} and
present simple analytical solutions in this regime.  We then
illustrate the utility of the kinetically dominated phase in
\Sref{sec:Kinetic_dominance_in_action} by application to a
spatially-flat Universe with polynomial and exponential potentials.
We enumerate the solutions that do not obey our broad assumptions in
\Sref{sec:When_is_kinetic_dominance_not_the_case?}.  We conclude in
\Sref{sec:Conclusions}.  Appendix~\ref{sec:uniqueness_theorem} proves a
uniqueness result crucial to the final step in the proof of kinetic
dominance.  

\section{Scalar field inflation models}
\label{sec:Scalar_field_inflation_models}
A universe comprised of multiple components  with densities
\(\{\rho_i\}\) and pressures \(\{P_i\}\) has the evolution equations:
\begin{align}
  \dot{H}+H^2 &= 
  -\frac{1}{6\m^2}\sum\limits_i\left( \rho_i + 3P_i\right), 
  \label{eqn:Raychaudhuri_rho}
  \\
  H^2 &= 
  \frac{1}{3\m^2}\sum\limits_i \rho_i, 
  \\
  \dot{\rho}_i 
  &= -3(\rho_i + P_i)H,  
  \label{eqn:conservation}
\end{align}
where \(H=\dot{a}/a\) is the Hubble parameter, \(a\) is the normalized
scale factor and a dot denotes differentiation with respect to cosmic
time, \(\dot{f}\equiv \d{f}/\d{t}\). The first equation is the {\em
acceleration\/} equation, and is derived from the trace of the
Einstein equations. The second is the {\em Friedmann\/} equation and
represents the conservation of energy. The third is the {\em
continuity\/} equation for the fluid \(\rho_i\). It should be noted that 
these equations are not independent, and that the acceleration
equation may be straightforwardly derived from the Friedmann and
continuity equations.  For convenience, we use Planck units
(\(G=c=\hbar=1\)) throughout, but for clarity retain the reduced Planck 
mass:
\[\m = \sqrt{\frac{\hbar c}{8\pi G}} = {(8\pi)}^{-1/2}.\]  

The simplest way to create a homogeneous and isotropic cosmological
background model which undergoes an inflationary phase is by assuming
that one of the fields is a real, time-dependent and homogeneous
scalar field \(\phi(t)\). The energy density and pressure of such a
field is given by:
\begin{equation}
  \rhoph = \tfrac{1}{2}\dot{\phi}^2+V(\phi),
  \qquad
  \Pph = \tfrac{1}{2}\dot{\phi}^2-V(\phi).
  \label{eqn:rhopdef}
\end{equation}
In addition to the scalar field, we shall allow the possibility of
including a collection of additional non-interacting fluids with
densities \(\{\rho_i\}\) and pressures \(\{P_i\}\) defined by their
equation-of-state parameters:
\begin{equation}
  w_i =\frac{P_i}{\rho_i},
  \label{eqn:equation_of_state}
\end{equation}
where \({w_i}\) are a set of constants determining the type of each
fluid. Some commonly assumed cosmological fluids are listed in
Table~\ref{tab:type_of_fluid} along with their \(w\)-values. Note that
we are accommodating the possibility of spatially curved universes
implicitly by including the case \(w_i=-1/3\). We shall term all of
these {\em auxiliary fluids}.
\begin{table}
  \caption{Commonly-assumed cosmological fluids and their
    equation-of-state parameters $w$, defined by equation
    \protect\eqref{eqn:equation_of_state}.  For more information on
    `missing matter', see Vazquez et al.\
    \protect\cite{vazquez_2012}\label{tab:type_of_fluid}
  }
  \begin{ruledtabular}
    \begin{tabular}{lc}
      Type of fluid & $w$ \\
      \hline
      Scalar field during KD & $ \phantom{-}1\phantom{/3} $ \\
      Radiation & $ \phantom{-}1/3 $ \\
      Matter & $ \phantom{-}0\phantom{/3} $ \\
      Spatial curvature & $ -1/3$ \\
      Missing matter & $ -2/3$ \\
      Dark energy (cosmological constant) & $ -1\phantom{/3} $ \\
    \end{tabular}
  \end{ruledtabular}
\end{table}

Using the notation in~\eqref{eqn:equation_of_state} and defining the
present-day densities \(\{\rho_{i,0}\}\), the evolution equations~\eqref{eqn:Raychaudhuri_rho}--\eqref{eqn:conservation} take
the form:
\begin{align}
  \dot{H}+H^2 &= 
  -\frac{1}{3\m^2}\left[\dot{\phi}^2 - V(\phi) +
  \sum_i\tfrac{1}{2}(1+3w_i)\rho_i\right] ,
  \label{eqn:Raychaudhuri_mod}
  \\
  H^2 &= 
  \frac{1}{3\m^2}\left[\tfrac{1}{2}\dot{\phi}^2 + V(\phi) +
  \sum_i\rho_i\right],
  \label{eqn:Friedmann_mod} 
  \\
  \rho_i &= 
  \rho_{i,0} \,a^{-3(1+w_i)},
  \label{eqn:rho_a} 
  \\ 
  0&= 
  \ddot{\phi} +3\dot{\phi}H + V^\prime(\phi).
  \label{eqn:Klein_Gordon_mod}
\end{align}

Inflation is defined as \(\ddot{a}>0\), or equivalently as
\(\dot{H}+H^2>0\). In the case when only an inflaton is present, this
condition can be recast in terms of the scalar field using the
acceleration equation~\eqref{eqn:Raychaudhuri_mod} as:
\begin{equation}
  \dot{\phi}^2<V(\phi).
  \label{eqn:Onset_inflation}
\end{equation}
The slow-roll inflation regime satisfies:
\begin{equation}
  \dot{\phi}^2\ll V(\phi).
  \label{eqn:Slow-roll}
\end{equation}
The amount of inflation is measured by the number of \(e\)-folds
\(N\propto \log a\), which is related to the Hubble parameter \(H\)
by:
\begin{equation}
  \dot{N}=H.\label{eqn:e-folds}
\end{equation}

For a generic potential \(V(\phi)\), there is no analytic solution for
the dynamics of a scalar field inflation model, even if no other
fluids are present. Hence, even in this simple case, the evolution
equations~\eqref{eqn:Raychaudhuri_mod}
and~\eqref{eqn:Klein_Gordon_mod} have to be integrated numerically using
suitable initial conditions at some time \(t=\ti\). In principle, 
\(t=\ti\) may be {\em any\/} cosmic time, although numerical stability of
the solution usually requires that the conditions are specified prior
to the onset of inflation.  Once any two of \(\phii \equiv \phi(\ti)\),
\(\dot{\phii} \equiv \dot{\phi}(\ti)\) and \(\Hi \equiv H(\ti)\) have been 
specified, the Friedmann equation~\eqref{eqn:Friedmann_mod} yields the
third. The quantities \(\Hi\), \(\phii\) and \(\dot{\phii}\) then provide
the necessary initial conditions for the integration of the coupled
dynamical equations~\eqref{eqn:Raychaudhuri_mod}
and~\eqref{eqn:Klein_Gordon_mod} for \(\phi(t)\) and \(H(t)\).

\section{Generic nature of kinetic dominance}
\label{sec:The_generic_nature_of_kinetic_dominance}

As has been previously observed \citep{Linde_initial_conditions_1985,
belinsky_inflationary_1985,particle_astrophysics_1990}, if one assumes
that at some point early in the Universe's history the opposite of the
slow roll condition~\eqref{eqn:Slow-roll} were true,
\begin{equation}
  \dot\phi^2\gg V(\phi),
  \label{eqn:kddef}
\end{equation}
then the evolution equations are analytically solvable.  We call this
condition kinetic dominance, since the potential energy of the field
\(V(\phi)\) is negligible in comparison to its kinetic energy
\(\frac{1}{2}\dot\phi^2\).

We shall restrict our attention to the very broad class of
cosmological models that satisfy:
\begin{equation}
|\dot{\phi}| > \vellim > 0 \qquad \text{as} \qquad a\to 0, 
\label{eqn:conditions}
\end{equation}
for some positive constant \(\vellim\).  This condition demands that
there be some epoch before which the inflaton evolves in a purely
monotonic manner, which we shall refer to as a {\em steadily moving
inflaton}. In this case, we find that the kinetic dominance
condition~\eqref{eqn:kddef} is entirely generic as \(a \to 0\), and holds
independently of the form of the potential \(V(\phi)\).

From~\eqref{eqn:kddef} it is then possible to show that the Universe
emerges from a singularity at a finite time in the past, which can be
set to \(t=0\). In addition, kinetic dominance also implies that the
(kinetic) energy density of the inflaton dominates the energy
densities of all of the other components \(\{\rho_i\}\) at early times, 
provided that \(w_i<1\). We shall leave the proof of these statements 
until \Sref{sec:consequences_of_kinetic_dominance}. 

We shall now prove the generic nature of kinetic dominance, i.e.\
that~\eqref{eqn:conditions} implies~\eqref{eqn:kddef}. The proof runs as 
follows:
\renewcommand{\theenumi}{\Alph{enumi}}
\begin{enumerate}
  \item                                        
    A new variable \(\Nflat\), termed the effective \(e\)-folds is
    introduced. This new variable enables one to assume without loss
    of generality that the potential \(V(\phi)\) is positive.
  \item
    The time coordinate \(t\) is re-scaled to a new time-like coordinate
    \(\tau\), termed {\em Halliwell\/} time. This removes the majority of
    the potential-dependence, and the two equations condense into a
    single equation in a new variable \(u\).
  \item
    The Hamilton-Jacobi representation is then utilised, exchanging
    Halliwell time for the field \(\phi\). The field \(\phi\) is then
    rescaled to a new variable \(\psi\), absorbing the Hubble parameter
    and implicitly all of the \(\{\rho_i\}\) dependence. One final
    monotonic transformation of the dependent variable \(u\) is made,
    leaving a single differential equation for a function \(y\) with
    \(\psi\) as the independent variable.
  \item
    The resulting equation has the property that, for any given
    potential \(V(\phi)\), there is at most a single solution \(f(\psi)\)
    that is both finite and positive. All other positive solutions
    \(y(\psi)\) are divergent.  When interpreted, a positive \(y(\psi)\)
    corresponds to a steadily moving inflaton, and a diverging
    \(y(\psi)\) represents a kinetically dominated universe.
\end{enumerate}

\subsection{Effective \(e\)-folds}
We define a new function \(\Hflat\) by the relation:
\begin{equation}
  \Hflat^2 = 
  \frac{1}{3\m^2}\left(\tfrac{1}{2}\dot{\phi}^2 + V_1+V(\phi) \right),
  \label{eqn:Friedmann_c} 
\end{equation}
where \(V_1\) is a positive constant, the value of which will be
discussed shortly. One may regard this as a modified Friedmann
equation, where the explicit dependence on the auxiliary fields has
been absorbed into a new parameter \(\Hflat\). The variable \(\Nflat\) can
be interpreted as the ``effective \(e\)-folds.'' In the case where the 
fluids may be neglected, one finds that \(\Hflat\approx\dot{N}\).

By differentiating the above definition, it is simple to show using
the Klein-Gordon equation~\eqref{eqn:Klein_Gordon_mod} that:
\begin{align}
  \frac{\Hflat \dHflat}{\dot{N}}  
  &=
  -\frac{\dot{\phi}^2}{2\m^2},
  \label{eqn:Raychaudhuri_c}
\end{align}
where \(\dot{N}=H\) from equation~\eqref{eqn:e-folds}. The Hubble
parameter \(\dot{N}\) can be related to the effective \(e\)-folds by
combining the Friedmann equation~\eqref{eqn:Friedmann_mod} with
equation~\eqref{eqn:Friedmann_c}:
\begin{equation}
  \dot{N}^2 = \Hflat^2 -\frac{V_1}{3\m^2} + \frac{1}{3\m^2}\sum_i\rho_i.
  \label{eqn:N_H_relation}
\end{equation}

Equations~\eqref{eqn:rho_a},~\eqref{eqn:Friedmann_c}--\eqref{eqn:N_H_relation} may
now be regarded as the evolution equations for the system in the
variables \(\{N,\Nflat,\phi,\rho_i\}\).  The potential \(V(\phi)\) now
only arises in~\eqref{eqn:Friedmann_c} in combination with \(V_1\).
Since all physical potentials are bounded below, one can choose \(V_1\)
such that \(V_1+V(\phi)\) is always positive. One can therefore treat
\(V_1+V(\phi)\) as the new ``effective'' potential and hence we
drop the \(V_1\) part from~\eqref{eqn:Friedmann_c} and assume \(V(\phi)\)
is positive.

\subsection{Halliwell time}
We now define a new time coordinate \(\tau\), such that:
\begin{equation}
  \frac{\d{\tau}}{\d{t}} 
  = 
  \sqrt{V(\phi)} \quad\Leftrightarrow\quad \tau 
  = 
  \int^t \sqrt{V(\phi)}\:\: \d{t}.
  \label{eqn:tau_def}
\end{equation}
This relation is well defined (up to a constant) and one-to-one as
from the above section one may assume that \(V(\phi)\) is finite and
positive. Physically,~\eqref{eqn:tau_def} corresponds to choosing a
measure of time in which the inflaton ``sees'' a near-constant
potential. This approach is analogous to the method used
by~\citet{halliwell_scalar_1987} in his work with exponential
potentials. We shall thus term this new timelike coordinate {\em
Halliwell time}.

\begin{figure}[t!]
  \includegraphics{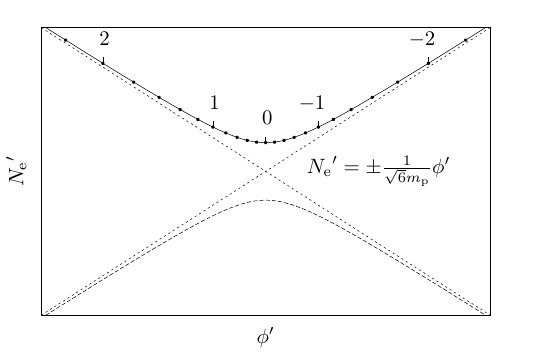}
  \caption{The constraint provided by the definition of \(\Nflat\) in
      Halliwell time. The dynamical variables \(\Nflatp\) and \(\prm \phi\) 
      lie on a hyperbola according to~\protect\eqref{eqn:Friedmann_tau}. A 
      natural parameterisation uses a hyperbolic angle \(u\) detailed
      in~\protect\eqref{eqn:Ntrans} and~\protect\eqref{eqn:phitrans}.
      Note that only the upper half is parametrized, as the lower half of the 
      hyperbola suggests a collapsing universe \((\Nflatp\propto
      H\propto\dot{a}<0)\). The points corresponding to
      \(u\in\{-2,-1,0,1,2\}\) have been labelled to guide the eye.}
      \label{fig:figure_hyperbola}
\end{figure}

Under this rescaling of time, the modified evolution
equations~\eqref{eqn:Friedmann_c} and~\eqref{eqn:Raychaudhuri_c} take the form:
\begin{align}
  \Nflatp^2 
  &= 
  \frac{1}{3\m^2}\left(\tfrac{1}{2}\prm{\phi}^2 + 1 \right),
  \label{eqn:Friedmann_tau} 
  \\
  -\frac{\prm{\phi}^2}{2\m^2} 
  &= 
  \frac{\Nflatp}{\prm{N}}\left(\Nflatpp 
  +\frac{1}{2} \prm{\phi}\Nflatp \frac{\d{}}{\d{\phi}}\log V \right),  
  \label{eqn:Raychaudhuri_tau}
\end{align}
where a prime denotes differentiation with respect to \(\tau\).
Equation~\eqref{eqn:Friedmann_tau} states that the dynamical variables
\(\Nflatp\) and \(\prm \phi\) lie on a hyperbola with asymptotic ratio
\({(\sqrt{6}\m)}^{-1}\), as illustrated in
\Fref{fig:figure_hyperbola}. Since one may take \(\Nflatp>0\), a
sensible parameterisation therefore is in terms of a hyperbolic angle
\(u\):
\begin{align}
  \Nflatp 
  &= 
  \frac{1}{\m\sqrt{3}}\cosh u,
  \label{eqn:Ntrans}
  \\
  \prm \phi 
  &= 
  -\sqrt{2}\sinh u.
  \label{eqn:phitrans}
\end{align}
Applying the transformation above to the Halliwell-time evolution
equations, we find that equation~\eqref{eqn:Friedmann_tau} is
trivially satisfied, and equation~\eqref{eqn:Raychaudhuri_tau} takes
the form:
\begin{equation}
  \frac{\Nflatp}{\prm{N}} \frac{\m}{\sqrt{6}}
  \left(\frac{\sqrt{2}}{\sinh u} \frac{\d{u}}{\d{\tau}} 
  - \frac{1}{\tanh u} \frac{\d{}}{\d{\phi}}\log V\right) 
  = -1.
  \label{eqn:master_tau}
\end{equation}

\subsection{Hamilton--Jacobi representation}
\label{sec:Hamilton-Jacobi_representations}
We reformulate the equation using the Hamilton--Jacobi representation;
instead of considering the variables as functions of time \(\tau\), one 
uses the field \(\phi\) as the independent variable. Since we are
considering universes with a monotonic inflaton \((\dot{\phi}\ne 0)\),
the transformation from \(t\) to \(\phi\) is monotonic, and hence so too 
is that from \(\tau\) to \(\phi\).

One can switch to the Hamilton--Jacobi representation by changing the
variables in the derivatives using the relation:
\begin{equation}
  \difrac{}{\tau}
  =
  \difrac{\phi}{\tau}\difrac{}{\phi}
  =
  \prm\phi\difrac{}{\phi}
  =
  -\sqrt{2}\sinh u \difrac{}{\phi},
  \label{eqn:tautrans}
\end{equation}
which on applying to equation~\eqref{eqn:master_tau} yields:
\begin{equation}
  \m\frac{\Nflatp}{\prm{N}} \sqrt{\frac{2}{3}}
  \left(\frac{\d{u}}{\d{\phi}} 
  + \frac{1}{2\tanh u} \frac{\d{}}{\d{\phi}}\log V\right) 
  = 
  1.
  \label{eqn:master_phi}
\end{equation}
We now  rescale the \(\phi\) field into a new field \(\psi\) via the
relation:
\begin{equation}
  \frac{\d{}}{\d{\psi}} 
  = 
  \m\frac{\Nflatp}{\prm{N}}\sqrt{\frac{2}{3}}\frac{\d{}}{\d{\phi}}.
\end{equation}
More explicitly, \(\psi\) is defined up to a constant by the monotonic
transformation:
\begin{equation}
  \frac{\d{\psi}}{\d{\phi}} 
  = 
  \sqrt{\frac{3}{2}}\frac{\prm{N}}{\m\Nflatp} 
  \quad
  \Leftrightarrow\quad \psi 
  = 
  \sqrt{\frac{3}{2}}\frac{1}{\m}\int^\phi 
  \frac{\prm{N}}{\Nflatp}\:d\phi,
  \label{eqn:psi_by_phi}
\end{equation}
and this relationship is well-defined since:
\begin{equation}
	\frac{\prm{N}}{\Nflatp} = \frac{H}{\Hflat} \ge 0.
\end{equation}
This rescaling absorbs all of the dependence on \(\prm N\) and thus
\(\{\rho_i\}\) via~\eqref{eqn:N_H_relation} into the definition of
\(\psi\). Under the transformation~\eqref{eqn:psi_by_phi}, the master
equation~\eqref{eqn:master_phi} takes the form:
\begin{equation}
  \frac{\d{u}}{\d{\psi}}  
  = 
  1-\frac{1}{\tanh u} \frac{\d{}}{\d{\psi}}\log \sqrt V.
\end{equation}
We have transformed the evolution
equations~\eqref{eqn:Raychaudhuri_mod}--\eqref{eqn:Klein_Gordon_mod} into a single 
master equation in one variable, where all of the potential
dependence is kept in a single term.

We may rearrange this slightly by making the monotonic transformation:
\begin{equation}
  u\mapsto y = \log\cosh(u), 
  \label{eqn:y_def}
\end{equation}
under which the master equation takes the form:
\begin{equation}
  \frac{\d{y}}{\d{\psi}} = \sqrt{1-e^{-2y}} - \frac{\d{}}{\d{\psi}}\log \sqrt V.
  \label{eqn:master_eq}
\end{equation}

\subsection{Interpreting the master equation}
\label{sec:interpreting_the_master_equation}

We now prove kinetic dominance by considering the asymptotics of the
master equation~\eqref{eqn:master_eq} in the limit \(a\rightarrow0\).  We 
prove that we may assume wlog that \(|\psi|\to\infty\) as \(a\to
0\). Given this, we prove that there is at most a single solution
\(f(\psi)\) which is finite, with all of the rest diverging as
\(a\rightarrow0\). We finish by showing that a diverging solution
\(y(\psi)\) of the master equation is equivalent to kinetic dominance.

We begin by examining the behaviour of \(\psi\) as \(a\rightarrow0\).  Through
elementary derivative transformations with the chain rule and the 
definitions of various variables, we find:
\begin{align}
  \frac{\prm\phi}{\Nflatp} 
  &=
  \frac{\prm\psi\frac{\d{\phi}}{\d{\psi}}}{\Nflatp},
  &\text{(chain rule)} 
  \nonumber
  \\
  &=
  \m\sqrt{\frac{2}{3}}\frac{\prm\psi}{\prm{N}},  
  &\text{(from equation \protect\ref{eqn:psi_by_phi})} 
  \nonumber
  \\
  &=
  \m\sqrt{\frac{2}{3}}\frac{\dot\psi}{\dot{N}},  
  &\text{(chain rule)} 
  \nonumber
  \\
  &=
  \m\sqrt{\frac{2}{3}}\frac{\d{\psi}}{\d{\log a}}.
  &\text{(since \(H=\frac{\d{}}{\d{t}}\log a\))}
  \label{eqn:div_psi}
\end{align}
Given the definition of Halliwell time~\eqref{eqn:tau_def}, on the
left hand side of the above expression, we have \(\prm\phi =
\dot\phi/\sqrt{V}\). Since \(\Nflatp>0\), and by
assumption~\eqref{eqn:conditions} \({|\dot\phi|>\xi>0}\) we thus find
that \(\psi\) is monotonic in \(a\).  The direction of the
monotonicity of \(\psi\) can be found by considering the
transformations we have made: %
\begin{equation}
  t
  \xrightarrow{\protect\eqref{eqn:tau_def}}
  \tau
  \xrightarrow{\protect\eqref{eqn:phitrans}}
  \phi
  \xrightarrow{\protect\eqref{eqn:psi_by_phi}}
  \psi.
\end{equation}
By considering the equations denoted above, one can see that
\(\frac{\d{\tau}}{\d{t}}>0\), \(\frac{\d{\psi}}{\d{\phi}}>0\), and
\(\frac{\d{\phi}}{\d{\tau}} = -\sqrt{2}\sinh u\). Thus as \(a\) and
\(t\) decrease, one finds that if \(u>0\), then \(\psi\) is
monotonically {\em increasing}.\footnote{This explains the choice of
    sign in the parametrization~\protect\eqref{eqn:phitrans}.}
As we are considering universes where \(\dot{\phi}\ne 0\), and wlog
\(V(\phi)>0\), this places the constraint that:
\begin{equation}
  u 
  = 
  \sinh^{-1}\left(\frac{\prm\phi}{\sqrt 2}\right) 
  = 
  \sinh^{-1}\left(\frac{\dot\phi}{\sqrt 2 V(\phi)}\right)
  \ne 
  0.
\end{equation}
The problem therefore breaks down into two possibilities: \(u>0\) and
\(\psi\) increasing, or \(u<0\) and \(\psi\) decreasing.  We will
consider the first possibility; the second may be treated in exactly
the same way, with a couple of sign changes.  

If \(\psi\) is monotonically increasing, then as \(a\to0\), either
\(\psi\to\psi_{\max{}}\) or it diverges \(\psi\to\infty\). In the
first of these possibilities, it follows that:
\begin{equation}
    \psi \to \psi_{\max{}}\quad \text{as}\quad a\to 0 \qquad \Rightarrow \qquad \frac{\d\psi}{\d\log a} \to 0.
\end{equation}
Now, since:
\begin{equation}
    \frac{\d\psi}{\d\log a} \propto \frac{\prm{\phi}}{\Nflatp} \propto \frac{\dot{\phi}}{\sqrt{\frac{1}{2}\dot{\phi}^2 + V(\phi)}},
\end{equation}
the assumption of a steadily moving inflaton~\eqref{eqn:conditions}
(i.e. \(|\dot\phi| > \xi > 0\) as \(a\to0\)) means that if  the left
hand side of the above tends to zero, then \(V(\phi)\) must diverge.
We now show that this set of statements contradict our initial
assumptions.  Integrating the master equation~\eqref{eqn:master_eq}
from some start point up to \(\psi_\smax\) yields:
\begin{equation}
  y(\psi) = \int^{\psi_\smax} \sqrt{1-e^{-2y}} \d{\psi} - \log\sqrt{V} + c.
\end{equation}
Given that the integrand on the right hand side is bounded, and that
the integral is over a finite range, the first term remains finite.
However, we know that when \(\psi\to\psi_{\max{}}\), the potential
\(V\) and hence \(\log\sqrt V\) must diverge. Thus, the solution \(y\)
in the above equation must become negative as \(\psi\to\psi_\smax\),
which contradicts the definition of \(y\) from~\eqref{eqn:y_def}.
Thus,  since \(\psi\not\to\psi_\smax\), we may assume that
\(\psi\to\infty\) as \(a\to0\).

We now show that as \(\psi\to\infty\) there is at most one {\em
finite\/} solution \(y\) satisfying the master
equation~\eqref{eqn:master_eq} while remaining non-zero.  Let us
assume that there exists a solution \(f(\psi)\)
of~\eqref{eqn:master_eq} that is positive and finite (\(0<f<\fm\) for
some finite \(\fm\)):
\begin{equation}
  \difrac{f}{\psi} = \sqrt{1-e^{-2f}} -\frac{\d{}}{\d{\psi}}\log \sqrt V,
  \label{eqn:master_eq_f}
\end{equation}
where:
\begin{equation}
  0<f(\psi)<\fm,
\end{equation}
and assume some initial condition on \(f\) at some finite value
\(\psi=\psiz\):
\begin{equation}
  f(\psiz)=\fz.
\end{equation}
Now consider a solution \(h(\psi)\) with some larger initial value:
\begin{equation}
  h(\psiz)=\hz>\fz.
\end{equation}
Since \(h\) is also a solution of the master equation~\eqref{eqn:master_eq} 
it satisfies:
\begin{equation}
  \difrac{h}{\psi} = \sqrt{1-e^{-2h}} -\frac{\d{}}{\d{\psi}}\log \sqrt V.
  \label{eqn:master_eq_h}
\end{equation}
Taking the difference of~\eqref{eqn:master_eq_f}
and~\eqref{eqn:master_eq_h} gives:
\begin{equation}
  \frac{\d{}}{\d{\psi}}\left(h-f\right)
  = 
  \sqrt{1-e^{-2h}}-\sqrt{1-e^{-2f}}.
  \label{eqn:master_eq_diff}
\end{equation}
By a uniqueness theorem, discussed in
Appendix~\ref{sec:uniqueness_theorem}, if \(h(\psiz)>f(\psiz)\), then
\(h(\psi_1)>f(\psi_1)\) (for \(\psi_1\ne\psi_0\)). Therefore one can see 
from~\eqref{eqn:master_eq_diff} that the difference between \(h\)
and \(f\) is monotonically increasing in any finite interval \([\psiz,\psi_1]\).  
One can thus conclude that:
\begin{equation}
  h-f>\hz-\fz=\Delta_0>0,
  \label{eqn:cosh_rel}
\end{equation}
where we have defined \(\Delta_0 = \hz-\fz\), and \(h\) and \(f\) are
evaluated at the end of the interval \(\psi_1\). Since \(h>\Delta_0+f\), 
it is easy to see using~\eqref{eqn:master_eq_diff} that:
\begin{equation}
  \frac{\d{}}{\d{\psi}}\left(h-f\right)
  > 
  \sqrt{1-e^{-2(f+\Delta_0)}}-\sqrt{1-e^{-2f}}.
\end{equation}
This is a monotonically decreasing function of \(f\), and hence
attains its minimum at \(\fm\); thus,
\begin{equation}
  \frac{\d{}}{\d{\psi}}\left(h-f\right)
  > 
  \sqrt{1-e^{-2(\fm+\Delta_0)}}-\sqrt{1-e^{-2\fm}} 
  >
  0.
\end{equation}
The difference between \(h\) and \(f\) is therefore monotonically
increasing at a rate greater than some positive number. Since \(f\) is
positive, it is bounded below and as \(\psi_1\) is made arbitrarily
large, \(h\) grows without bound.

\begin{figure}[t!]
  \includegraphics{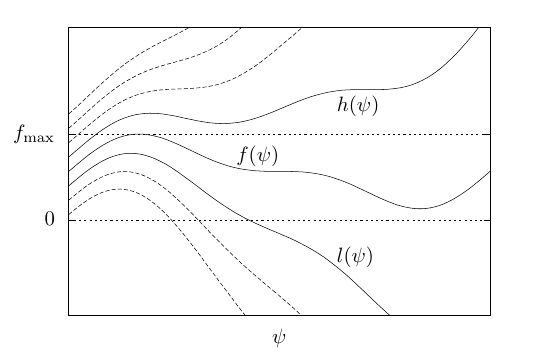}
  \caption{In general there is at most a single solution \(f(\psi)\) to 
    the master equation~\protect\eqref{eqn:master_eq} that is positive
    and finite \(0<f(\psi)<\fm\). Any solution \(h(\psi)\) that begins
    higher than \(f(\psi)\) must diverge. Consequently, any solution
    \(l(\psi)\) that begins lower than \(f(\psi)\) cannot remain above
    \(0\). If it were able to, then \(l(\psi)\) would also be another
    finite solution lying between \(0\) and some \(l_\smax\). By the
    previous argument, this would mean \(f\) could not be finite,
    contradicting our initial assumption.}
    \label{fig:figure_uschem}
\end{figure}
We therefore find ourselves in a situation demonstrated in
\Fref{fig:figure_uschem}. For any given potential \(V(\phi)\), there is 
at most a single solution \(f(\psi)\) that is finite and positive
[\(0<f(\psi)<\fm\)]: any solution that is larger than \(f(\psi)\) at some 
point \(\psi=\psiz\) diverges as \(\psi\to\infty\) (\(a\to0\)).  Further, any 
solution \(l(\psi)\) that starts out less than \(f(\psi)\) must fall to a 
value less than \(0\).  If \(l(\psi)\) did not fall below \(0\), but were 
another example of a finite positive solution, then by the argument 
above, this would imply \(f(\psi)>l(\psi)\) must diverge, contradicting 
our initial assumptions on \(f(\psi)\).

One therefore expects universes with a steadily moving inflaton to
have a generically diverging \(y\) as \(a\to0\), except perhaps for a
single special case for a given potential \(V(\phi)\).  The consequences 
of a generically divergent \(y\) shall now be examined.  If \(y\) 
diverges, then so does \(u\) by equation~\eqref{eqn:y_def}.  If \(u\) 
diverges, then we find that \(\prm \phi\) diverges
by~\eqref{eqn:phitrans}:
\begin{equation}
  \frac{\d{\phi}}{\d{\tau}}=\prm{\phi} 
  = 
  -\sqrt{2}\sinh u\rightarrow -\infty.
\end{equation}
Converting back from Halliwell time to cosmic time using
equation~\eqref{eqn:tau_def} shows:
\begin{equation}
  \prm{\phi}^2 
  = 
  \frac{\dot{\phi}^2}{V(\phi)}.
\end{equation}
One can thus see that the divergence of \(y\), \(u\) and \(\prm \phi\)
therefore requires that:
\begin{equation}
  \lim\limits_{y\to\pm\infty} \frac{\dot\phi^2}{V(\phi)}
  =
  \lim\limits_{a\to 0} \frac{\dot\phi^2}{V(\phi)} = \infty,
  \label{eqn:kdfinal}
\end{equation}
which is equivalent to saying that \(\dot{\phi}^2\gg V(\phi)\) as
\(a\to0\). The early universe is generically kinetically dominated.

\section{Consequences of kinetic dominance}
\label{sec:consequences_of_kinetic_dominance}

The condition \(\dot\phi^2\gg V(\phi)\) for kinetic dominance allows one 
to derive several results. First, the kinetic energy of the inflaton
dominates over the other fluids as \(a\to 0\), allowing us at early
times to neglect any additional effects such as curvature, radiation,
matter or a cosmological constant. Second, the universe emerges from
an initial singularity at a finite coordinate time, which may be taken
as \(t=0\). Finally, one is able to determine exact analytic expressions for 
the solutions in coordinate or conformal time for each of the 
cases where curvature, radiation, matter or a cosmological constant 
are present.

\subsection{Dominance of \(\dot{\phi}^2\) over other fluids}
\label{sec:dominance_fluids}
In the limit that \(a\to 0\), the auxiliary fluid with the largest value 
of \(w\) dominates over all of the others. Along with kinetic dominance, 
we can therefore assume that the acceleration~\eqref{eqn:Raychaudhuri_mod} 
and Friedmann~\eqref{eqn:Friedmann_mod} 
equations take the form:
\begin{align}
  \dot{H}+H^2 
  &= 
  -\frac{\dot{\phi}^2}{3\m^2} - \frac{1}{6\m^2}(1+3w)\rho_w,
  \label{eqn:Ray_KD_eq_rho}
  \\
  H^2 
  &= 
  \frac{\dot{\phi}^2}{6\m^2} +  \frac{1}{3\m^2}\rho_w,
  \label{eqn:Friedmann_KD_eq_rho}
\end{align}
where \(\rho_w\) is the density of the auxiliary fluid with the largest 
\(w\).  It is not difficult to show using equation~\eqref{eqn:rho_a}, 
along with \(H=\frac{\d{}}{\d{t}}\log a\) that these equations solve to give:
\begin{align}
  H^2 
  &= 
  \frac{1}{3\m^2}\left(\frac{\beta^2}{a^6} + \rho_w\right),
  \label{eqn:H_of_a_1}
  \\
  \dot{\phi}^2 
  &= 
  2\frac{\beta^2}{a^6},
  \label{eqn:dotphi_of_a_1}
\end{align}
where \(\beta\) is an integration constant. From this, if \(w<1\) then as
\(a\to 0\), one finds:
\begin{equation}
  \dot{\phi}^2 \propto a^{-6} 
  \gg
  \rho_w \propto a^{-3(1+w)}.
  \label{eqn:dotphidom}
\end{equation}
Thus, the kinetic term of the inflaton dominates over all other fluids
with \(w<1\).

Historically, inflationary potentials were considered in the context of 
grand unified theories \citep{PhysRevLett.48.1220,linde_1982} which
resulted in an effective potential \(V(\phi,T)\) depending on the value 
of the field \(\phi\) and a temperature.  
This was then developed 
\citep{1995PhRvL..74.1912B,PhysRevLett.75.3218} into a theory in which 
the inflaton remains in thermal equilibrium with an auxiliary 
radiation fluid.

More recent work \citep{2007PhRvD..76f3512P} typically assumes that
the inflaton is decoupled from the auxiliary fluids in the
preinflationary phase, and that the Universe is {\em radiation
dominated\/} at this early stage. Given the above result~\eqref{eqn:dotphidom}, 
such assumptions may now need revisiting.

\subsection{Finite time singularity}

Since \(H = \dot{a}/a\), we can express the coordinate time \(t\) as an 
integral: 
\begin{equation}
  t = \int \frac{\d{a}}{aH}.
\end{equation}
From equation~\eqref{eqn:H_of_a_1}, one can see that \({(aH)}^{-1}\) is 
finite as \(a\to0\). By the above integral, this shows that the Universe 
emerges at a finite time in the past, which can be taken as \(t=0\).  
Moreover, from~\eqref{eqn:dotphidom}, the (dominant) energy density of 
the Universe scales as \(a^{-6}\) as \(a \to 0\), showing that \(t=0\) is a 
singularity.

\subsection{Analytic solutions for the kinetically dominated universe} 
If one considers the solutions of the acceleration and Friedmann
equations~\eqref{eqn:Raychaudhuri_mod} and~\eqref{eqn:Friedmann_mod}
in the limit that \(a\to0\), then one can neglect the potential term
\(V(\phi)\) as it is suppressed by \(\dot{\phi}^2\). In addition, the
other fluid terms are negligible in comparison to the term with the
largest \(w\).  When these considerations are taken into account, the
evolution equations take the form shown in~\eqref{eqn:Ray_KD_eq_rho}
and~\eqref{eqn:Friedmann_KD_eq_rho}. One can find solutions for
\(\phi(a),H(a)\) and \(\rho_w(a)\) parametrically in terms of \(a\).
In addition, one can find coordinate time \(t(a)\) in terms of \(a\)
using the relation:
\begin{equation}
  t = \int \frac{\d{a}}{aH(a)}.
\end{equation}
Conformal time \(\eta\) is defined by the equation \(\dot{\eta} =
a^{-1}\), and can be found in terms of \(a\) using:
\begin{equation}
  \eta = \int \frac{\d{a}}{a^2H(a)}.
\end{equation}
The solutions are:
\begin{align}
  \rho_w(a) 
  &\propto 
  a^{-3(1+w)} ,
  \\
  H{(a)}^2 
  &= 
  \frac{1}{3\m^2}\left(\frac{\beta^2}{a^6} + \rho_w\right)  ,
  \label{eqn:H_of_a}
  \\
  \dot{\phi}{(a)}^2
  &=
  2\frac{\beta^2}{a^6} ,
  \label{eqn:dotphi_of_a}
  \\
  \phi(a)
  &=
  c  \pm
  \sqrt{\frac{2}{3}}\frac{\m}{1-w} \log \left[ 
  \frac 
  {{a}^{3(1-w)}}
  {{\left(\sqrt{1+\frac{\rho_w{a}^{6}}{\beta^2}}+1\right)}^2} 
  \right]  ,
  \label{eqn:phi_of_a}
  \\
  t(a)
  &=
  a^3 \frac{\m\sqrt{3}}{3\beta} 
  \:\: 
  {_2F_1}\left(
  \frac{1}{2},
  \frac{1}{1-w};
  \frac{2-w}{1-w},
  -\frac{a^6\rho_w}{\beta^2}
  \right) ,
  \label{eqn:t_of_a}
  \\
  \eta(a) 
  &= 
  a^2 \frac{\m\sqrt{3}}{2\beta}
  \:\: 
  {_2F_1}\left(
  \frac{1}{2},
  \frac{2}{3(1-w)};
  \frac{5-3w}{3(1-w)},
  -\frac{a^6\rho_w}{\beta^2}
  \right),
  \label{eqn:eta_of_a}
\end{align}
where \(\beta\) and \(c\) are constants of integration which will be
redefined shortly. We have chosen \(t,\eta\) such that \(a\to0\) as
\(t,\eta \to 0\).

For specific values of \(w\), the hypergeometric functions \(_2F_1\) take 
simple forms. If \(w=-1\) or \(0\), then equation~\eqref{eqn:t_of_a} is 
expressible in closed form, in terms of trigonometric and algebraic 
functions in \(a\) respectively. If \(w=-1/3\) or \(1/3\), then~\eqref{eqn:eta_of_a} 
may be expressed in closed form. In each of these 
cases, these equations are invertible giving an expression for \(a(t)\) 
or \(a(\eta)\).  We note that, except for the case \(w=-1/3\), the 
solutions~\eqref{eqn:H_of_a}--\eqref{eqn:eta_of_a} correspond to a 
spatially-flat universe.

We shall examine each of the above cases in turn, after first looking
at the case in which there are no auxiliary fields, \(\rho_w=0\).  
In so doing, it will prove useful to define the functions:
\begin{equation}
  \begin{array}{rr}
    S_k(x),&S_k^{-1}(x)
    \\ 
    C_k(x),&C_k^{-1}(x)
    \\ 
    T_k(x),&T_k^{-1}(x)
    \\
  \end{array}
  =
  \left\{
  \begin{array}{rl}
    \begin{array}{rr}
      \sin(x),&\arcsin(x)
      \\ 
      \cos(x),&\arccos(x)
      \\ 
      \tan(x),&\arctan(x)
      \\
    \end{array}
    &: k>0 \\
    \begin{array}{rr}
      x,&x
      \\ 
      1,&1
      \\ 
      x,&x
      \\
    \end{array}
    &: k=0 \\
    \begin{array}{rr}
      \sinh(x),&\arcsinh(x)
      \\ 
      \cosh(x),&\arccosh(x)
      \\ 
      \tanh(x),&\arctanh(x)
      \\
    \end{array}
    &: k<0
  \end{array}
  \right. 
\end{equation}

\subsubsection{No auxiliary fields, \(\rho_w=0\)}
If \(\rho_w=0\), then equation~\eqref{eqn:t_of_a} becomes:
\begin{equation}
  t = a^3 \frac{\m\sqrt{3}}{3\beta}.
\end{equation}
One can rearrange this to find \(a\) as a function of \(t\),
\begin{equation}
  a(t)
  =
  t^{1/3} {\left(\frac{3\beta}{\m\sqrt{3}}\right)}^{1/3},
  \label{eqn:a_of_t_beta_flat}
\end{equation}
and then substitute this into equations~\eqref{eqn:H_of_a}
and~\eqref{eqn:dotphi_of_a} to find:
\begin{align}
  H(t) 
  &= 
  \frac{1}{3t}, 
  \label{eqn:H_of_t_flat} 
  \\
  \dot{\phi}(t) 
  &= 
  \pm\sqrt{\frac{2}{3}}\frac{\m}{t}.
  \label{eqn:dotphi_of_t_flat}
\end{align}
The latter integrates to give:
\begin{equation}
  \phi(t) 
  = 
  \phip \pm\sqrt{\frac{2}{3}}\m\log\left(\frac{t}{\tp}\right),  
  \label{eqn:phi_of_t_flat}
\end{equation}
where \(\phip\) is an integration constant chosen such that
\(\phi(\tp)=\phip\), where \(\tp\) is some time. It is more appropriate to 
re-define the integration constant \(\beta\) as:
\begin{equation}
  \beta 
  \equiv 
  \frac{\Rp^3\m}{\tp\sqrt{3}}, 
  \label{eqn:beta_def}
\end{equation}
since then, equation~\eqref{eqn:a_of_t_beta_flat} for \(a(t)\) becomes:
\begin{equation}
  a(t) 
  = 
  \Rp {\left(\frac{t}{\tp}\right)}^{1/3},
  \label{eqn:a_of_t_flat}
\end{equation}
which is more in keeping with equation~\eqref{eqn:phi_of_t_flat}.  

For this case, one can also obtain analytical solutions in terms of
conformal time.  If \(\rho_w=0\) and \(\beta\) is defined by~\eqref{eqn:beta_def}, 
equation~\eqref{eqn:eta_of_a} becomes:
\begin{equation}
  \eta(t) 
  = 
  a^2 \frac{\m\sqrt{3}}{2\beta}
  =
  \frac{3\tp}{2\Rp^3}a^2.
  \label{eqn:eta_of_t}
\end{equation}
Using equation~\eqref{eqn:a_of_t_flat}, we can show:
\begin{equation}
  \eta
  =
  \etap{\left(\frac{t}{\tp}\right)}^{2/3},
\end{equation}
where we have defined \(\etap\) as:
\begin{equation}
  \etap
  =
  \frac{3\tp}{2\Rp}.
  \label{eqn:etap_eq}
\end{equation}
Now we have \(\eta(t)\) in equation~\eqref{eqn:eta_of_t}, we can change
equations~\eqref{eqn:H_of_t_flat},~\eqref{eqn:dotphi_of_t_flat},~\eqref{eqn:phi_of_t_flat}
and~\eqref{eqn:a_of_t_flat} to:
\begin{align}
  a(\eta)
  &=
  \Rp {\left(\frac{\eta}{\etap}\right)}^{1/2},
  \\
  H(\eta)
  &=
  \frac{1}{3\tp}{\left(\frac{\eta}{\etap}\right)}^{-3/2},
  \\
  \dot{\phi}(\eta)
  &=
  \pm\sqrt{\frac{2}{3}}
  \frac{\m}{\tp}{\left(\frac{\eta}{\etap}\right)}^{-3/2},
  \\
  \phi(\eta)
  &=
  \phip \pm\sqrt{\frac{3}{2}}\m\log\left(\frac{\eta}{\etap}\right). 
  \label{eqn:phi_flat}
\end{align}

It should be noted that since \(\dot{\phi}^2 \gg \rho_w\) at
sufficiently early times, all solutions reduce to the above forms for
small enough \(t\) or \(\eta\). We can thus fix the form of solutions
with nonzero \(\rho_w\) by matching onto the above solutions for
sufficiently small \(t\) or \(\eta\).  

\subsubsection{Dark energy, \(w=-1\)}
For dark energy in the form of a cosmological constant, we find that
the energy density in standard notation is:
\begin{equation}
  \rho_w = \m^2\Lambda.
\end{equation}
For \(w=-1\), equation~\eqref{eqn:t_of_a} is expressible in terms of
trigonometric functions, and may be rearranged to express the scale
factor \(a\) in terms of coordinate time. Once \(a(t)\) is obtained, the 
remaining equations~\eqref{eqn:H_of_a},~\eqref{eqn:dotphi_of_a} and~\eqref{eqn:phi_of_a} 
can be used to express the rest of the variables 
in terms of \(t\). Using our definition of \(\beta\) (equation \nolinebreak\ref{eqn:beta_def}), and defining the new timescale,
\begin{equation}
  \tL = \frac{1}{\sqrt{3\Lambda}},
  \label{eqn:tl}
\end{equation}
the solutions are:
\begin{align}
  a(t)
  &=
  \Rp{\left[{\frac {S_{-\Lambda}(t/\tL)}{\tp/\tL}}\right]}^{1/3},
  \\
  H(t)
  &=
  \frac{1}{3\tL}\frac{1}{T_{-\Lambda}(t/\tL)},
  \\
  \dot{\phi}(t)
  &=
  \sqrt{\frac{2}{3}}\frac{\m}{\tL}\frac{1}{S_{-\Lambda}(t/\tL)},
  \\
  \phi(t)
  &=
  \phip \pm \sqrt{\frac{2}{3}}\m
  \log\left[
  \frac{\tL}{\tp} 
  \frac 
  {2\: S_{-\Lambda} \left( t/\tL \right) }
  {1 + C_{-\Lambda} \left( t/\tL \right)}  
  \right].
\end{align}

\subsubsection{Spatial curvature, \(w=-1/3\)}
Spatial curvature is equivalent to a fluid with equation-of-state
parameter \(w=-1/3\) and density:
\begin{equation}
  \rho_w = -3\m^2\frac{\kappa}{a^{2}}.
\end{equation}
For \(w=-1/3\), equation~\eqref{eqn:eta_of_a} is expressible in terms of 
trigonometric functions, and may be rearranged to express the scale
factor \(a\) in terms of conformal time. Once \(a(\eta)\) is obtained, the 
remaining equations~\eqref{eqn:H_of_a},~\eqref{eqn:dotphi_of_a}
and~\eqref{eqn:phi_of_a} can be used to express the rest of the variables 
in terms of \(\eta\). Using our definitions of \(\beta\)
(equation~\ref{eqn:beta_def}) and \(\etap\) 
(equation~\ref{eqn:etap_eq}), and defining the new timescale,
\begin{equation}
  \etak = \frac{1}{2\sqrt{\kappa}},
  \label{eqn:etak}
\end{equation}
the solutions are:
\begin{align}
  a(\eta)
  &=
  \Rp{\left[
  \frac
  {S_\kappa\left(\eta/\etak\right)}
  {\etap/\etak} \right]}^{1/2},
  \\
  H(\eta)
  &=
  \frac{1}{3\tp}
  \frac{{(\etap/\etak)}^{3/2}}
  {T_\kappa(\eta/\etak)\sqrt{S_\kappa(\eta/\etak)}}, 
  \\
  \dot{\phi}(\eta)
  &=
  \pm\sqrt{\frac{2}{3}}
  \frac{\m}{\tp}
  {\left[
  \frac
  {\etap/\etak}
  {S_{\kappa}(\eta/\etak)}
  \right]}^{3/2},
  \\
  \phi(\eta) 
  &=
  \phip \pm \sqrt{\frac{3}{2}}\m\log  \left[
  \frac{\etak}{\etap} 
  \frac{2\:S_{\kappa}\left(\eta/\etak \right) }
  {1 + C_{\kappa} \left( \eta/\etak \right)   }  
  \right]. 
\end{align}

\subsubsection{Matter, \(w=0\)}
For matter with zero pressure, one has \(w=0\) and so:
\begin{equation}
  \rho_w = \rhomp{\left(\frac{a}{\Rp}\right)}^{-3},
\end{equation}
where \(\rhomp\) is an integration constant, labelling the energy density 
of matter at the epoch \(\Rp\).  For \(w=0\), equation~\eqref{eqn:t_of_a} 
is expressible as an algebraic function, and may be rearranged to 
express the scale factor \(a\) in terms of coordinate time. 
Once \(a(t)\) is obtained, the remaining equations~\eqref{eqn:H_of_a},~\eqref{eqn:dotphi_of_a}~\&~\eqref{eqn:phi_of_a}
can be used to express 
the rest of the variables in terms of \(t\).  Using our definition of 
\(\beta\) (equation~\ref{eqn:beta_def}), and defining the new timescale,
\begin{equation}
  \tm = \frac{4\m^2}{3\tp\rhomp },
  \label{eqn:tm}
\end{equation}
the solutions are:
\begin{align}
  a(t)
  &=
  \Rp{\left(\frac{t}{\tp}\right)}^{1/3}
  {\left(1+\frac{t}{\tm}\right)}^{1/3} ,
  \\
  H(t) &= 
  \frac{1+2\frac{t}{\tm}}{3t{{\left( 1+\frac{t}{\tm} \right) }}},
  \\
  \dot{\phi}(t) &= 
  \pm\sqrt{\frac{2}{3}}\m
  \frac{1}
  {t \left( 1+\frac{t}{\tm} \right) },
  \\
  \phi(t) &=
  \phip \pm \sqrt{\frac{2}{3}}\m \log\left[  
  \left(\frac{t}{\tp}\right) 
  \frac {1}{1+\frac{t}{\tm}}  
  \right].
\end{align}

\subsubsection{Radiation, \(w=1/3\)}

For radiation one has \(w=1/3\), and so:
\begin{equation}
  \rho_w = \rhorp{\left(\frac{a}{\Rp}\right)}^{-4},
\end{equation}
where \(\rhorp\) is an integration constant, labelling the energy density 
of matter at the epoch \(\Rp\).  For \(w=1/3\), equation~\eqref{eqn:eta_of_a} 
is expressible as an algebraic function, and may 
be rearranged to express the scale factor \(a\) in terms of coordinate 
time. Once \(a(\eta)\) is obtained, the remaining equations~\eqref{eqn:H_of_a},~\eqref{eqn:dotphi_of_a} and~\eqref{eqn:phi_of_a} can 
be used to express the rest of the variables in terms of \(\eta\). Using 
our definition of \(\beta\)~\eqref{eqn:beta_def}, and \(\etap\)~\eqref{eqn:etap_eq}, and
defining the new timescale:
\begin{equation}
  \etar = \frac{3\m^2}{\Rp^2\etap\rhomp},
  \label{eqn:etar}
\end{equation}
the solutions are:
\begin{align}
  a(\eta)
  &=
  \Rp{\left(\frac{\eta}{\etap}\right)}^{1/2}
  {\left({1+\frac{\eta}{\etar}}\right)}^{1/2},
  \\
  H(\eta) 
  &= 
  \frac{1}{3\tp}{\left(\frac{\eta}{\etap}\right)}^{-3/2}
  \frac{1+ 2\frac{\eta}{\etar}}
  {{\left(1+ \frac{\eta}{\etar}\right)}^{3/2}},
  \\
  \dot{\phi}(\eta) 
  &=
  \sqrt{\frac{2}{3}}\frac{\m}{\tp}
  {\left(\frac{\eta}{\etap}\right)}^{-3/2}
  \frac{1}{{\left(1+ \frac{\eta}{\etar}\right)}^{3/2}},
  \\ 
  \phi(\eta) 
  &=
  \phip+\sqrt {\frac{3}{2}}\m\log  
  \left[
  \left(\frac{\eta}{\etap}\right)
  \frac {1}{\left(1+ \frac{\eta}{\etar}\right)} 
  \right].
\end{align}

\subsection{The constants of integration}
\label{sec:constants}
In the previous section, several constants arose, which we shall now
review. For the system of equations~\eqref{eqn:Ray_KD_eq_rho}--\eqref{eqn:Friedmann_KD_eq_rho},
one would expect four constants of 
integration. The first is chosen by setting \(a=0\) at \(t=0\). For the 
case \(\rho_w=0\), the second and third are chosen by choosing a later 
time \(\tp>0\) and fixing:
\begin{align}
  \phi(\tp) &= \phip \nonumber,\\
  a(\tp) &= \Rp \nonumber.
\end{align}
We can also determine conformal time in this case, which involved
defining a new constant \(\etap\) in terms of \(\Rp\) and \(\phip\) via 
equation~\eqref{eqn:etap_eq}, and the solutions may be determined in 
conformal time. For the remaining cases, there is an additional
integration constant determined by the value of \(\rho_w\) when the
scale factor is \(\Rp\). The solutions for these cases are then
determined by matching them onto the case \(\rho_w=0\) at early times.

In addition, we defined a relevant time scale for each of the \(\rho_w
\neq 0\) cases: \(\tL\), \(\etak\), \(\tm\), \(\etar\). These are expressed in terms of 
the previous integration constants in equations~\eqref{eqn:tl},~\eqref{eqn:etak},~\eqref{eqn:tm}
and~\eqref{eqn:etar}. If one chooses 
\(\tp\) or \(\etap\) to be much less than this second time scale, then the 
Universe is in a fully kinetically dominated regime at \(\tp\), \(\etap\), 
with solutions very close to the \(\rho_w=0\) case. When \(\eta\) is of 
the order of the second time scale, then the effects of the Universe's 
(dominant) additional component can be seen.

Although we have determined these equations in terms of three
integration constants, there are in fact only two.  The evolution
equations~\eqref{eqn:Raychaudhuri_mod}~\&~\eqref{eqn:Friedmann_mod}
possess a two-parameter symmetry corresponding to a re-scaling of \(a\)
and \(t\). More precisely, the form of the equations does not change
under the transformation:
\begin{equation}
  a\mapsto\alpha a, 
  \qquad 
  t \mapsto\sigma^{-1}t,
\end{equation}
provided that \(\rho_i\) and the potential transform as:
\begin{equation}
  \rho_i \mapsto \alpha^{3(1+w_i)}\sigma^2\rho_i, 
  \qquad
  V(\phi) \mapsto \sigma^2 V(\phi).
\end{equation}
This symmetry can be used effectively to remove some of the remaining
integration constants. In practice this means that one may set \(\tp=1\) 
to be the Planck time and the scale factor \(a(\tp)=\Rp=1\).  Usually one 
takes the scale factor to be unity at the present epoch \(a_0\equiv a(t_0)=1\), 
but this requirement is complicated by the uncertainties of 
reheating, so we do not follow that convention here.  One may
physically interpret \(\phip\) as the value of the field at \(t=t_p\).  
This requires extrapolating the classical equations far beyond their 
validity, so it is more of a mnemonic aid than a physical interpretation.  
As we will see below, \(\phip\) controls the total number of \(e\)-folds 
of inflation.

\section{Kinetic dominance in action}
\label{sec:Kinetic_dominance_in_action}

We shall now demonstrate the utility of kinetic initial conditions in
the analysis of inflationary models. Even without integrating the
evolution equations for \(H(t)\) and \(\phi(t)\), one sees that the basic 
scenario entails the universe emerging from an initial singularity at
\(t=0\) in a regime where the kinetic energy of the inflaton dominates
its potential energy along with any curvature or additional fluids.
The evolution of \(H(t)\) and \(\phi(t)\) in this regime are given
by~\eqref{eqn:H_of_t_flat} and~\eqref{eqn:phi_of_t_flat} at sufficiently 
early times.  \(H(t)\) and \(|\phi(t)|\) and their time derivatives
decrease during this period of kinetic dominance, which concludes when
there is approximate equipartition \(\dot{\phi}^2 \sim V(\phi)\) between 
the kinetic and potential energies of the inflaton.  This marks the 
onset of a (typically brief) period of fast-roll inflation
\citep{Linde:2001}, which must eventually become slow-roll inflation,
with \(\dot{\phi}^2 \ll V(\phi)\), since the latter is a generic
attractor solution for inflation models
\citep{belinsky_inflationary_1985}. We will see that integration of 
the equations of motion in some illustrative cases does indeed verify 
these expectations.

For simplicity we shall work in the case with no other additional
fields, \(\rho_w=0\), although our methods apply equally well to more
complicated solutions. After discussing the validity of the initial
conditions and numerical techniques, we shall consider two forms of
potential: polynomial and exponential.

\subsection{Initial conditions and scaling}

For \(\rho_w=0\) the Universe is spatially-flat and contains only the
inflaton field. The evolution equations then take the form:
\begin{align}
  H^2 
  &= 
  \frac{1}{3\m^2}
  \left(\frac{1}{2}\dot{\phi}^2 + V(\phi)\right),
  \label{eqn:Friedmann_eq} 
  \\
  0
  &= 
  \ddot{\phi} +3\dot{\phi}H + V^\prime(\phi).
  \label{eqn:Motion_eq_Scalar}
\end{align}
The general solution to the evolution equations has the asymptotic
form given in equations~\eqref{eqn:H_of_t_flat}
and~\eqref{eqn:phi_of_t_flat}. As discussed in \Sref{sec:constants}, we may 
choose \(\tp=1\) to be the Planck time. Given this, we set the initial 
conditions at an initial time \(\ti\) as:
\begin{align}
  \phi(\ti) \equiv \phii
  &= 
  \phip - \sqrt{\frac{2}{3}}\m\log \ti, 
  \label{eqn:phi_ic}
  \\
  \dot\phi(\ti) 
  \equiv 
  \phidi
  &= 
  -\sqrt{\frac{2}{3}}\frac{\m}{\ti}, 
  \label{eqn:phid_ic}
  \\
  H(\ti) 
  \equiv 
  \Hi
  &= 
  \frac{1}{3\ti}. 
  \label{eqn:H_ic}
\end{align}
There is a single constant of integration \(\phip\), which directly
controls the number of \(e\)-folds during inflation. The number of
\(e\)-folds \(N_*\) between the pivot scale \(k_*\) exiting the Hubble radius and the end of inflation 
is typically \(50\)--\(60\) \citep{planck_collaboration_planck_2013-1}. For 
the rest of this paper \(\phip\) will be chosen so that the total number 
of \(e\)-folds \(N_\mathrm{tot}=65\). This will be discussed in greater
detail in \Sref{sec:powspec}.

Throughout this paper we work in the classical regime. In order for
the conditions above to be valid, the initial conditions must be set
at a time greater than the Planck time, \(\ti >\tp=1\), but within the
kinetic dominated regime, for which \(V(\phii) \ll \phidi^2\). Setting
\(\ti=\tp=1\) in the above, one sees that kinetic dominance will endure 
beyond the Planck time provided:
\begin{equation}
  V(\phip) \ll \m^2.
\end{equation}
The above requirement typically holds for potentials that give
physically reasonable inflation models. For example, in the case of a
free inflaton with mass \(m\), one has \[ V(\phi) = \frac{1}{2}m^2
\phi^2.\] In order to generate the correct amplitude of curvature
perturbations, the mass must be of the order \(m\sim10^{-5}\m\), whereas 
to generate the correct number of \(e\)-folds one requires
\(\phip\bigO{10}\), in which case \(V(\phip) \sim 10^{-8} 
\m^2\).  Thus, there is no need to advocate trans-Planckian physics, 
since kinetic dominance lasts well beyond the Planck time, so one can 
set \(\ti \gg \tp\).

We note that the evolution equations~\eqref{eqn:Friedmann_eq}~\&~\eqref{eqn:Motion_eq_Scalar} 
are invariant under  the simultaneous 
re-scaling of the time coordinate, Hubble parameter and inflaton 
potential:
\begin{align}
  t 
  &\mapsto 
  \sigma^{-1}t,
  \\
  H 
  &\mapsto 
  \sigma H,
  \\
  V(\phi) 
  &\mapsto
  \sigma^2 V(\phi).
\end{align}
The advantage of this for numerical work is that a multiplicative
scaling parameter from the potentials can be removed without loss of
generality.

\subsection{Polynomial potentials}
\label{sec:section_polynomial_potentials}
We begin by analysing examples of polynomial potentials of the form:
\begin{equation}
  \Vpol(\phi) = \mu^2\phi^n.
  \label{eqn:polpot}
\end{equation}
To obtain results we integrate the evolution equations~\eqref{eqn:Friedmann_eq}~\&~\eqref{eqn:Motion_eq_Scalar} numerically.  
Our kinetic initial conditions are chosen using equation~\eqref{eqn:phi_ic}--\eqref{eqn:H_ic} 
with an initial time \(\ti\) small 
enough such that the inflaton is in the kinetic regime
\(V(\phii)\ll\dot{\phii}^2\). For the purposes of numerics the scaling
parameter \(\mu\) can be removed by rescaling the time coordinate
(setting \(\sigma=\mu^{-1}\)). \(\phip\) is set by requiring that there 
be \(55\) \(e\)-folds during inflation.

\begin{figure}[t!]
  \includegraphics{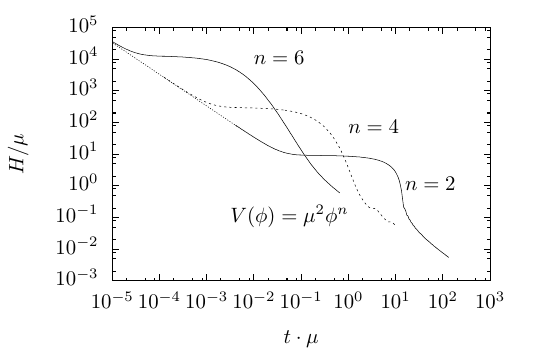}
  \caption{The evolution of the Hubble parameter for three polynomial
    potentials of the form in~\eqref{eqn:polpot}. The axes have been
    rescaled in terms of the parameter \(\mu\) in the potential, so that 
    this graph describes the evolution for any choice of \(\mu\).  The 
    initial conditions for inflation were set using the flat-universe 
    kinetic conditions, and the parameter \(\phip\) was chosen so as to 
    give \(55\) \(e\)-folds of inflation. All three universes emerge in a 
    kinetically dominated phase with \(H=1/(3t)\), before entering a 
    slow-roll inflationary phase with \(H\sim\mathrm{constant}\). The 
    universe then exits inflation, after which small `wiggles' in \(H\) 
    can be seen. These are due to the field \(\phi\) executing 
    oscillations about the base of the potential.}
    \label{fig:figure_Hpol}
\end{figure}
\begin{figure}[t!]
  \includegraphics{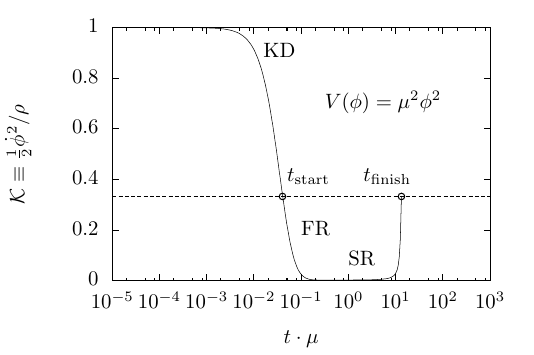} 
  \caption{The evolution of \(\Kfrac = \frac{1}{2}\dot{\phi}^2/\rho\)
  for the quadratic inflaton potential \(V(\phi) = \mu^2 \phi^2\).  The 
  Universe can be seen to begin in a kinetically dominated state (KD).  
  It enters inflation at \(t_\mathrm{start}\). There is a brief
  fast-roll (FR) phase of inflation, before the Universe enters a
  protracted slow-roll (SR) phase. The Universe exits inflation at
  time \(t_\mathrm{finish}\) after \(55\) \(e\)-folds. After the end of 
  inflation the field \(\phi\) executes oscillations about the base of 
  the potential. This causes \(\Kfrac\) to oscillate rapidly between \(0\) 
  and \(1\). For clarity, the value of \(\Kfrac(t)\) with 
  \(t>t_\mathrm{finish}\) has not been plotted. The above behaviour is 
  common to all of the  polynomial potentials shown in
  \protect\Fref{fig:figure_Hpol}. }
  \label{fig:figure_Kpol}
\end{figure}

The evolution of the Hubble parameter is shown in
\Fref{fig:figure_Hpol} for polynomials with \(n=2,4,6\). It is helpful
to define the variable:
\begin{equation}
  \Kfrac 
  \equiv  
  \frac{\frac{1}{2}\dot{\phi}^2  }{\rho}  
  =  
  \frac{\frac{1}{2}\dot{\phi}^2  }
  {\frac{1}{2}\dot{\phi}^2 + V(\phi)},
\end{equation}
to be used as an investigative tool. \(\Kfrac\) is the ratio of the
kinetic energy to the total energy and has the properties that:
\begin{equation}
  \Kfrac 
    \left\{
    \begin{array}{rl}
      \approx 1 &\Rightarrow \hbox{kinetic dominance} 
      \\
      >\frac{1}{3} &\Rightarrow \hbox{not inflating}
      \\
      <\frac{1}{3} &\Rightarrow \hbox{fast-roll/power-law inflation}
      \\
      \approx 0 &\Rightarrow \hbox{slow-roll inflation.}
    \end{array}
    \right.
\end{equation}
This is used as a diagnostic tool in \Fref{fig:figure_Kpol} for the
quadratic potential \(V(\phi) = \mu^2\phi^2\). Examining
\Fref{fig:figure_Kpol} one can see that our earlier expectations are
verified. There are four stages of evolution:
\begin{enumerate}
  \item the Universe emerges from an initial singularity in a
    kinetically dominated phase,
  \item it transitions through fast-roll inflation, \label{itm:fr}
  \item before entering a protracted slow-roll phase, 
  \item and thereafter the field \(\phi\) quickly moves towards a
      minimum of the potential, about which it executes a decaying
      oscillation.
\end{enumerate}
The fast-roll transition in point~\eqref{itm:fr} is potentially
responsible for the damping of the CMB spectrum at low-\(\ell\) observed 
in recent cosmological data. This will be discussed more fully in 
\Sref{sec:powspec}.

\subsection{Exponential potentials}
\label{sec:Exponential_potentials}
We now consider inflaton potentials of the form:
\begin{equation}
  \Vpow(\phi) 
  = 
  2V_0\left[
  \cosh\left(\frac{\sqrt{2\epsilon}}{\m}\phi\right)-1
  \right],
  \label{eqn:coshpot}
\end{equation}
which is a symmetrised form of the more common exponential potential;
as \(\phi\rightarrow\pm\infty\) the potential takes the asymptotic form:
\begin{equation}
  V(\phi) 
  = 
  V_0 \exp\left(\frac{\sqrt{2\epsilon}}{\m} |\phi|\right).
  \label{eqn:exp_pot}
\end{equation}
Exponential potentials~\eqref{eqn:exp_pot} have been well studied
\citep{yokoyama_dynamics_1988}. For potentials of this form, the
evolution equations have the analytical power-law solutions:
\begin{align}
  a(t) 
  &\propto 
  {t}^{1/\epsilon},
  \label{eqn:pow_law_a_sol}
  \\
  \phi(t)
  &=
  \pm\m\sqrt{\frac{2}{\epsilon}}
  \log\left(\sqrt{\frac{V_0}{\left(3-\epsilon\right)}}
  \frac{\epsilon}{\m} t\right),
  \\
  H(t)
  &=
  \frac{1}{\epsilon t}.  
  \label{eqn:pow_law_H_sol}
\end{align}
It is worth noting that for mathematical consistency one requires
\(\epsilon < 3\). For \(\epsilon<1\) these solutions are (continuously) 
inflating and are thus termed `power-law 
inflation'~\citep{lucchin_power-law_1985}. Moreover, these are 
attractor solutions as the universe evolves forwards in time. It is
also straightforward to show that at all epochs \(\dot{\phi}^2/V(\phi)
= 2\epsilon/(3-\epsilon)\), so the ratio of the inflaton kinetic energy 
to its potential energy is constant. In particular, one notes that the 
solution is kinetically dominated only in the limit \(\epsilon \to 3\).  
We may also interpret \(\epsilon\) as the slow-roll parameter:
\begin{equation}
  \epsilon\equiv\epsilon_H = -\frac{\dot{H}}{H^2},
\end{equation}
although one does not need to assume that it is small.

At first sight,
solutions~\eqref{eqn:pow_law_a_sol}--\eqref{eqn:pow_law_H_sol} appear to represent 
a counter-example to kinetic dominance as \(t \to 0\), so it is worth exploring them further 
in the context of our proof of the generic 
nature of kinetic dominance in 
\Sref{sec:The_generic_nature_of_kinetic_dominance}. 

In terms of the master equation~\eqref{eqn:master_eq}, in a flat
universe, one finds \(\phi = \sqrt{\frac{2}{3}}\m \psi + \mathrm{const.} \) 
and thus for the exponential potential~\eqref{eqn:exp_pot}:
\begin{equation}
  \frac{\d{}}{\d{\psi}} \log \Vpow = \frac{2\epsilon}{\sqrt{3}}. 
\end{equation}
Consequently, the master equation has the constant, finite solution:
\begin{equation}
  f(\psi) = \log{\left(1-\frac{4}{3}\epsilon^2\right)}^{-1/2}.
  \label{eqn:uf_power_law}
\end{equation}
However, from the proof in
\Sref{sec:The_generic_nature_of_kinetic_dominance}, we know that this
finite solution is unique. Any solution which is greater than this
diverges as \(|\phi|\to\infty\), and any solution less than this becomes 
negative.  Indeed, this is already evident from the fact that the 
power-law solutions are attractors as the universe evolves forwards 
in time. By the same token, these solutions are unstable in the limit \(t 
\to 0\), i.e.\ travelling backwards in time one diverges away from 
these solutions and generically arrives at kinetic dominance or a
turn-around.  We note that the proof that the power-law solutions are
attractors is due to \citet{halliwell_scalar_1987}, and our work in
\Sref{sec:The_generic_nature_of_kinetic_dominance} demonstrates that
Halliwell's methodology is applicable more generally.

\begin{figure}[t!]
  \includegraphics{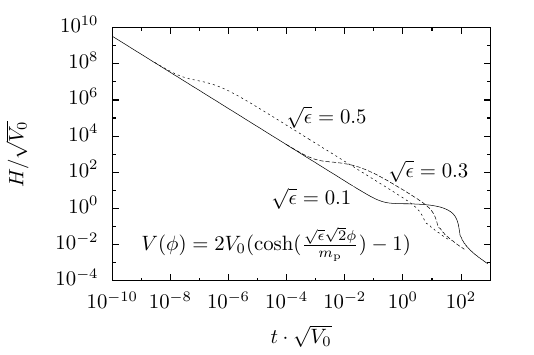}
  \caption{As in Figure~\protect\ref{fig:figure_Hpol}, but for the
  hyperbolic cosine potential~\protect\eqref{eqn:coshpot}, which tends
  to an exponential potential for \(\phi \to \pm\infty\). The scaling
  constant is now \(\sqrt{V_0}\), and three values of \(\sqrt{\epsilon}\) 
  are considered. One can see clearly that the Universe emerges in a
  kinetically dominated state with \(H=1/(3t)\).  The Universe then
  enters a protracted power-law phase where \(H = 1/(\epsilon t)\).  The 
  field \(\phi\) oscillates about the base of the potential after the 
  exit of the inflationary phase, causing wiggles in the later
  sections of the above plot. }
  \label{fig:figure_Hlam}
\end{figure}
\begin{figure}[t!]
  \includegraphics{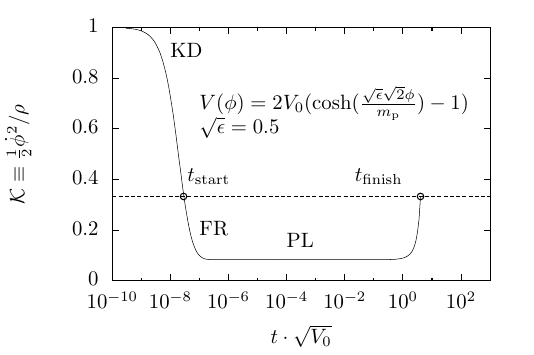}
  \caption{As in Figure~\protect\ref{fig:figure_Kpol}, but for the
  hyperbolic cosine inflaton potential with \(\sqrt{\epsilon}=0.5\).  The 
  Universe emerges in a kinetically dominated (KD) phase before going
  through transitory fast-roll (FR). Instead of a slow-roll inflation,
  the Universe then settles into a power-law inflation (PL),
  characterised by \(\Kfrac=\epsilon/3\). The Universe exits inflation
  at \(t_\mathrm{finish}\), after which \(\phi\) executes oscillations
  about the base of the potential. \(\Kfrac\) therefore oscillates
  rapidly between \(0\) and \(1\), and the later \(t>t_\mathrm{finish}\) 
  section of this plot has been suppressed for clarity. }
  \label{fig:figure_Klam}
\end{figure}

We show the evolution of the Universe governed by the hyperbolic
cosine inflaton potential~\eqref{eqn:coshpot} in
Figures~\ref{fig:figure_Hlam}~\&~\ref{fig:figure_Klam}. The analysis
is the same as that presented in \Sref{sec:section_polynomial_potentials}.  
As in our previous example, the Universe emerges from the initial
singularity in kinetic dominance, which then transitions through a
brief period of fast-roll inflation into a generically long-lasting
power-law inflation state until the exit is reached as
\(\phi\rightarrow0\), which corresponds to the minimum of the potential.  

\subsection{Another example of a finite solution \(f(\psi)\)}
The \(f(\psi)\) described above in equation~\eqref{eqn:uf_power_law} is 
one of the simplest examples of a finite solution. As shown earlier,
there is at most one such finite solution for any given potential. For
concreteness we demonstrate another less trivial example in this
section.

\begin{figure}[t!]
  \centering
  \includegraphics{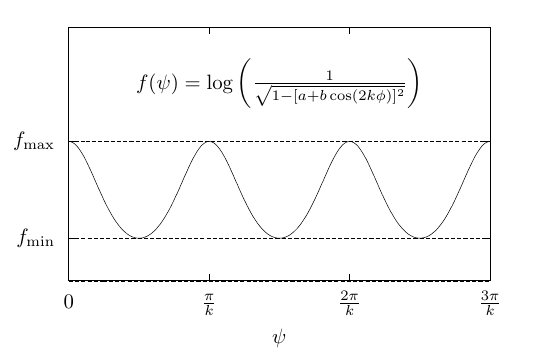}
  \caption{An oscillating finite solution to the master equation~\protect\eqref{eqn:master_eq} 
  with \(V(\psi)\) defined as in equation~\protect\eqref{eqn:Vphi_uf_example} 
  and demonstrated in 
  \protect\Fref{fig:figure_ospot}. If \(a\) and \(b\) are chosen such 
  that \(a>b>0\), \(0<a+b<1\), \(0<a-b<1\), then this represents a finite, 
  positive solution.  From equation~\protect\eqref{eqn:uf_example} it 
  is easy to see that \(\fmn = \log\left(\frac{1}{\sqrt{1-{(a+b)}^2}}\right)\),
      \(\fm=\log\left(\frac{1}{\sqrt{1-{(a-b)}^2}}\right)\). }
      \label{fig:figure_uf}
\end{figure}

\begin{figure}[t!]
  \centering
  \includegraphics{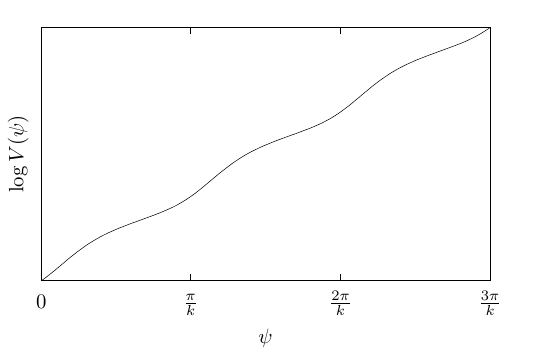}
  \caption{The master equation may be solved for the potential
      \(V(\psi)\) if a form of \(f(\psi)\) is chosen. For the choice of 
      \(f(\psi)\) detailed in equation (\protect\ref{eqn:uf_example}), the 
      potential is solvable in closed form (equation \protect\ref{eqn:Vphi_uf_example}).
  }
  \label{fig:figure_ospot}
\end{figure}

By reverse-engineering the master equation~\eqref{eqn:master_eq_f}, one 
can find a potential \(V(\psi)\) for any specified \(f(\psi)\).  For
example, if one chooses the oscillating solution (shown in
\Fref{fig:figure_uf}):
\begin{equation}
  f(\psi) = \log\left( \frac{1}{\sqrt{1-{[a+b\cos(2k\psi)]}^2}}\right),
  \label{eqn:uf_example}
\end{equation}
then this \(f(\psi)\) is the finite solution of the potential defined by:
\begin{equation}
	V(\psi)
    =
    \left[ 1-{(a+b \cos2k\psi)}^2 \right]
    e^{2 a \psi +\frac{b}{k} \sin 2k\psi},
    \label{eqn:Vphi_uf_example}
\end{equation}
whose shape is detailed in \Fref{fig:figure_ospot}.

To show explicitly that this is the only finite solution in this case, we consider a perturbed solution \(y(\psi) = f(\psi)+\delta(\psi)\).  From the uniqueness theorem (Appendix~\ref{sec:uniqueness_theorem}) if \(\delta\) is initially positive (negative), then it is positive (negative) for all \(\psi\). From the master equation~\eqref{eqn:master_eq} one may show that the perturbed solution satisfies:
\begin{align}
  \frac{\d{}}{\d{\psi}}\delta 
  =& 
  \sqrt{1-e^{-2(f+\delta)}} -\sqrt{1-e^{-2f}} 
  \nonumber
  \\
  &>
  \sqrt{1-e^{-2(\fm+\delta)}} -\sqrt{1-e^{-2\fm}},
  \label{eqn:perturb}
\end{align}
where \(\fm = \log(1/\sqrt{1-{(a-b)}^2})\). Substituting this in, one finds:
\begin{align}
  \frac{\d{}}{\d{\psi}}\delta 
  > 
  \sqrt{1-e^{-\delta}({1-{(a-b)}^2})} - \sqrt{1-({1-{(a-b)}^2})}.
\end{align}
One can see that if \(\delta>0\), then the right hand side is strictly
greater than some number greater than zero, hence \(\delta\) grows
without bound, and any solution greater than \(f\) initially diverges. 

Working from equation~\eqref{eqn:perturb}, one also finds:
\begin{align}
  \frac{\d{}}{\d{\psi}}\delta 
  <&
  \sqrt{1-e^{-2(\fmn+\delta)}} -\sqrt{1-e^{-2\fmn}} 
  \nonumber
  \\
  &<
  \sqrt{1-e^{-\delta}({1-{(a+b)}^2})} - \sqrt{1-({1-{(a+b)}^2})}.
\end{align}
One can see that if \(\delta<0\), then the right-hand side is strictly
less than some number less than zero, hence \(\delta\) falls without
bound, and any solution less than \(f\) eventually becomes negative. 

Thus one finds that \(f(\psi)\) is the only finite positive solution.
All other solutions either become negative or diverge.


\subsection{Power spectrum of the curvature perturbation}
\label{sec:powspec}

The most interesting aspect of kinetic dominance is seen when the
power spectrum of scalar curvature perturbations is examined. Recent
observations of CMB power spectra
\citep{hinshaw_nine-year_2012,planck_collaboration_planck_2013} show
an unexpected suppression at low multipoles. Whilst these deviations
are not large enough to cause us to discard the standard \(\Lambda\)
cold dark matter (\(\Lambda\)CDM) cosmology
\citep{1998PhRvD..57.2207B,2000PhRvD..62l3513B,2004PhRvD..69f3516D},
there is still enough tension to be worthy of investigation.  As we
show below, kinetic dominance predicts a generic cutoff in the
curvature power spectrum at large spatial scales. This is precisely
what is needed to suppress low multipole moments of the CMB power
spectrum whilst retaining the quality of the fit at higher \(\ell\)
values.

\begin{figure}[t!]
  \centering
  \includegraphics[width=\columnwidth]{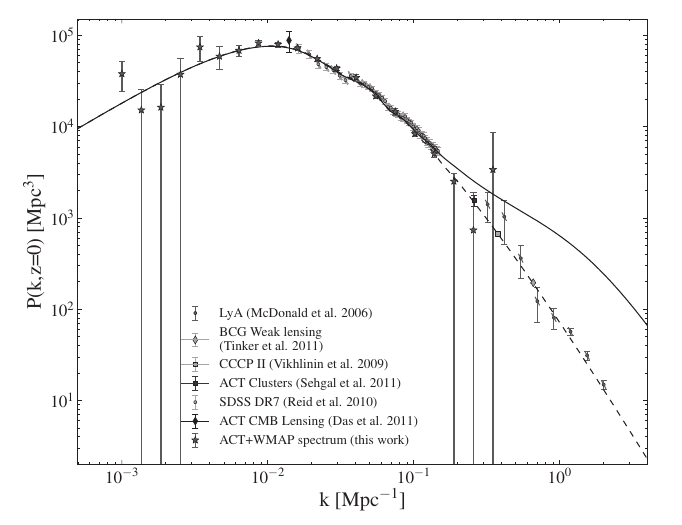}
  \caption{Figure taken from \protect\citet{hlozek_atacama_2012}
  showing the current status of the observed late-time matter
  perturbation power spectrum. As one can see, the currently probed
  \(k\) range is \(10^{-3}<k<2\:\mathrm{Mpc}^{-1}\).   }
  \label{fig:experimental_power_spectrum}
\end{figure}

\Fref{fig:experimental_power_spectrum} has been taken from 
\citet{hlozek_atacama_2012} and shows the current status of the
observational constraints on the late-time matter power spectrum,
given by:
\begin{equation}
    \Punnorm(k,z=0) = 2\pi^2 k \Pnorm_\mathcal{R}(k) G^2(z) T^2(k),
\end{equation}
where \(G(z)\) gives the growth of matter perturbations, \(T(k)\) is the 
matter transfer function and \(\PR(k)\) is the 
primordial curvature perturbation power spectrum. This mapping enables 
one to combine constraints on the power spectrum from CMB and other 
probes at \(z\approx 0\).

As shown by \citet{liddle_cosmological_2000}, the primordial curvature
perturbation power spectrum \(\PR(k)\) is given 
approximately by:
\begin{equation}
  \PR(k)
  =
  {\left(\frac{H^2}{2\pi\dot{\phi}}\right)}^2_{k=a H},
  \label{eqn:curvature_power_spectrum}
\end{equation}
where, as denoted, the right-hand side is evaluated when a given scale
crosses the horizon. If one has numerically calculated \(a(t)\), \(H(t)\) and 
\(\dot{\phi}(t)\), then plotting \({\left(H^2/ 2\pi\dot{\phi}\right)}^2\) 
against \(aH\) will give the shape of the spectrum.

In order to perform predictive calculations, one must calibrate the
\(aH\) axis to an observable scale today. This is easy to do if one
defines a comoving pivot scale \(k_*\), which leaves the horizon (at a
time \(t_*\)) when \(N_*\) \(e\)-folds of inflation remain. In general, the 
relation between \(k_*\) and \(N_*\) depends on both the potential 
\(V(\phi)\) and the details of cosmic reheating. For most reasonable 
models, \(50<N_*<60\)  for \(k_*\) with a value of
\(0.05\:\mathrm{Mpc}^{-1}\) today
\citep{planck_collaboration_planck_2013-1}. For this work, we will
take \(N_*=55\). 

\begin{figure}[t!]
  \includegraphics{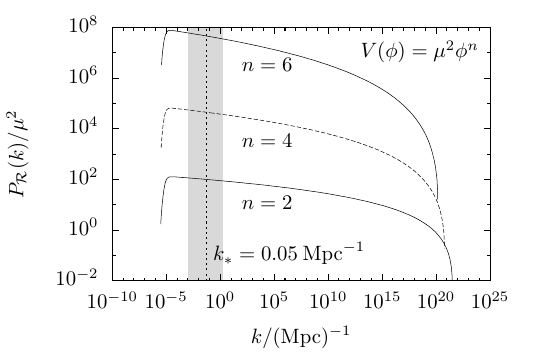}
  \caption{The approximate power spectrum of the primordial curvature
      perturbations for polynomial potentials, calculated using
      equation~\protect\eqref{eqn:curvature_power_spectrum}. The units of the \(k\) 
      axis are determined by the requirement that there are \(N_*=55\) 
      \(e\)-folds remaining when the pivot scale
      \(k_*=0.05\:\mathrm{Mpc}^{-1}\) exits the horizon \(k=aH\). The
      magnitude of the power spectrum is determined by the scaling \(\mu\) 
      in the potential, and the low-\(k\) cutoff is determined by a
      choice of \(\phip\) such that there are \(N_\mathrm{tot}=65\)
      \(e\)-folds of total inflation. The grey area indicates the angular scales that 
      have been experimentally probed; see
      \protect\Fref{fig:experimental_power_spectrum}. }
      \label{fig:figure_CSpol}
\end{figure}

\begin{figure}[t!]
  \includegraphics{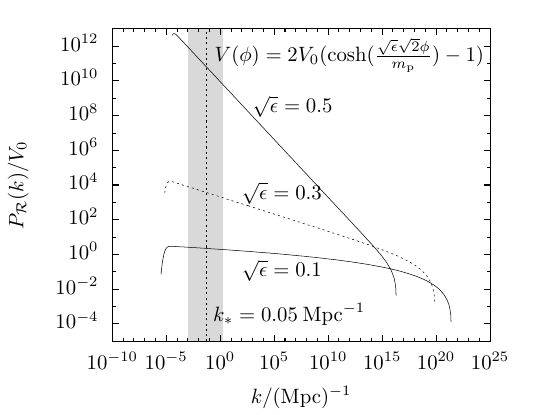}
  \caption{As in Figure~\protect\ref{fig:figure_CSpol}, but for exponential 
  potentials. The magnitude of the power spectrum now scales with 
  \(V_0\) rather than \(\mu^2\).  }
  \label{fig:figure_CSlam}
\end{figure}

Once a value for \(N_*\) is chosen, one can determine numerically the
time \(t_*\) at which \(N_*\) \(e\)-folds of inflation remain as well as 
\(a_*\equiv a(t_*)\) and \(H_*\equiv H(t_*)\). Since we know that the 
value of \(aH\) at \(t_*\) corresponds to a wave number today of
\(0.05\:\mathrm{Mpc}^{-1}\), we may calibrate the \(aH\) axis of the plot 
of the power spectrum using:
\begin{equation}
  k_\mathrm{today} 
  = 
  \frac{aH}{a_*H_*}\times0.05\:\mathrm{Mpc}^{-1}.
\end{equation}
Calibrated plots of \(\PR(k)\) are found in
Figures~\ref{fig:figure_CSpol} and~\ref{fig:figure_CSlam}.

The shape of the primordial power spectra obtained corresponds to that
found in \citep{lasenby_closed_2003}. We see that, in general,
\(\PR(k)\) has less power at low and high \(k\) values 
than would be expected from a canonical power-law primordial spectrum.
The low-\(k\) cutoff is entirely generic and occurs as a result of the
brief period of fast-roll prior to slow-roll or power-law inflation.
This effect has been discussed previously
by \cite{boyanovsky_cmb_2006}. The fast-roll regime 
behaves like an attractor potential in the wave equations for the mode
functions of curvature and tensor perturbations. This potential leads
then to the suppression of the primordial power spectra at low \(k\).
Hence, it might be able to account for the suppression of the
quadrupole of the CMB in agreement with observational data, as
discussed in \citep{boyanovsky_cmb_2006-1}.  The exact position of the
low-\(k\) cutoff is determined by the value of \(\phip\), as it controls 
the total number of \(e\)-folds of inflation. This effect has also been 
discussed in the context of ``Open Inflation''
\citep{Yamauchi_strings_2011,Linde_open_1999,Linde_toy_1999}, and
examined using WMAP data in \citet{Contaldi_suppress_2003}.

Further inspection shows \(\PR \sim \log k\) after the low-\(k\) cut off. 
This is identical to the result found by
\citet{lasenby_closed_2003} and in contrast with the standard
power-spectrum parameterisation which assumes a near-flat power-law
scaling: \(\log\PR\sim \log k\).

\begin{figure}[t!]
  \centerline{\includegraphics{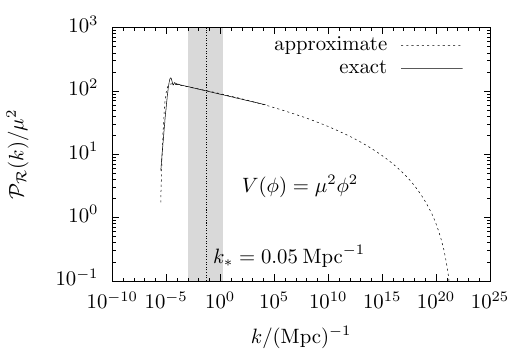}}
  \caption{As in Figure~\protect\ref{fig:figure_CSpol}, but now
  comparing the approximate and exact calculations of the primordial
  power spectrum for an \(n=2\) polynomial potential. The key feature
  of a low \(k\) suppression remains, with an additional ringing
  effect. The details of this ringing effect depend strongly on how
  one chooses initial conditions for the comoving curvature
  perturbation in the kinetically dominated phase. For this
  calculation, we chose Bunch Davies initial conditions. }
  \label{fig:new_CSpol}
\end{figure}

It should be observed that we are using the approximation~\eqref{eqn:curvature_power_spectrum} 
outside the slow-roll regime for 
which it is valid; nonetheless we have performed full calculations
that do not use the above approximation and which indicate that the
resulting power spectrum is, in fact, a good representation of the
true spectrum. These approximate spectra demonstrate the key generic
aspects of the accurate calculation: both exhibit a low-\(k\) cut off
and that \(\PR(k) \sim \log(k)\).  We shall follow 
this work with a second publication containing the full details and 
discussion of the accurate calculation, but a representative example
is shown in \Fref{fig:new_CSpol}. It should be noted that such
accurate calculations depend strongly on how one chooses initial
conditions in the kinetically dominated phase for the comoving
curvature perturbation. An alternative but related 
accurate calculation has been performed by \citep{Lello_tensor_2013},
which uses kinetic initial conditions to show that the suppression at
low-\(\ell\) is entirely generic. It should also be noted that methods
which reconstruct the primordial power spectrum \(\PR(k)\)
\citep{vazquez_reconstruction,Hazra_reconstruction_2013} using data
also show a dip at low \(k\) values.

\begin{figure*}[t!]
  \centerline{\includegraphics{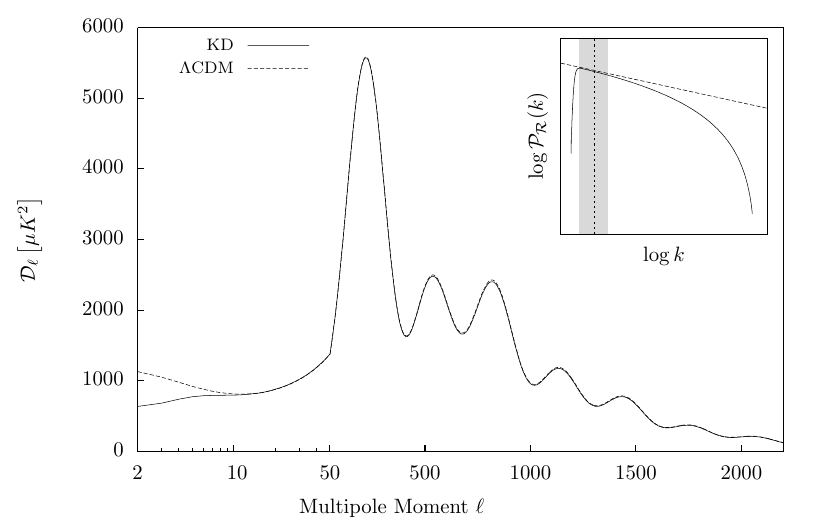}}
  \centerline{\includegraphics{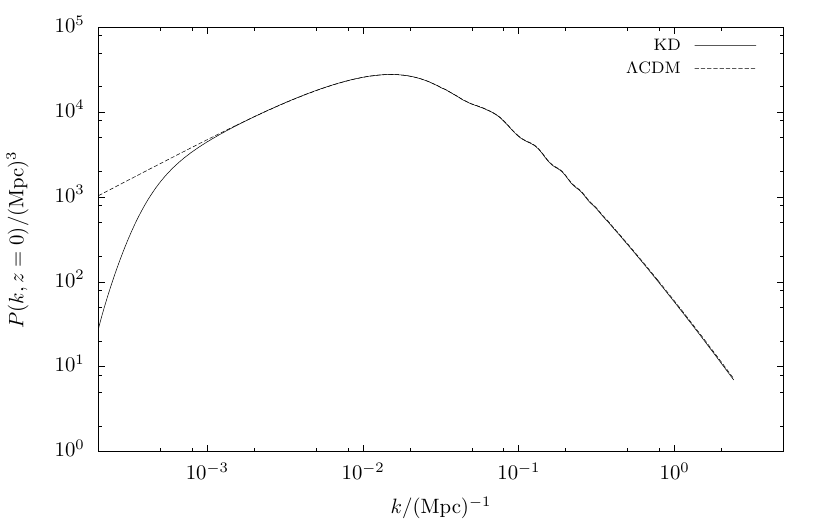}}
  \caption{The CMB scalar power spectra (top) and the late-time matter
      power spectra (bottom), resulting from the curvature perturbation 
      power spectra in the inset of the top figure. The solid lines
      corresponds to a free inflaton with potential
      \(V=\frac{1}{2}m^2\phi^2\) assuming kinetic initial conditions with 
      \(m=0.81\times10^{-5}\m\), \(\phip = 21.8\), \(N_*=42.5\).  The dashed 
      lines correspond to the best-fit standard \(\Lambda\)CDM model.  The 
      presence of the cut off in the curvature power spectrum for 
      kinetic initial conditions causes a suppression of power on large 
      angular scales in both the CMB and matter power spectra. }
      \label{fig:figure_Cl}
\end{figure*}

We demonstrate the suppression on large angular scales of the CMB and
late-time matter power spectra qualitatively in \Fref{fig:figure_Cl}.
In the standard six-parameter \(\Lambda\)CDM cosmology, the primordial
power spectrum has a power-law form, parametrized by two variables
\(A_s\) and \(n_\mathrm{s}\), such that:
\begin{equation}
  \PR(k) = A_s{\left(\frac{k}{k_*}\right)}^{n_\mathrm{s}-1}.
\end{equation}
Using the best-fit parameters from {\em Planck\/}+WP+highL+BAO
\citep{planck_collaboration_planck_2013} yields the standard matter
and CMB power spectra (dashed lines), whereas the alternative power
spectra (solid line) were generated using \(\PR(k)\) 
from \Fref{fig:figure_CSpol} (for \(n=2\)), for which the axes were
rescaled to agree with the values of \(A_s\) and \(n_\mathrm{s}\) during 
the slow-roll phase. The resulting matter and CMB power spectra are 
seen to exhibit a suppression of power at low \(k\) and \(\ell\) values, 
with the rest of the spectra perfectly intact, as required by
cosmological observations.  Further investigation is clearly required,
but this analysis already demonstrates the utility of kinetic
dominance.


The results of this section agree with the ``just enough inflation''
scenario introduced by  Ramirez {\it et al.\ } 
\citep{Ramirez_excluded_2009,Ramirez_predictions_2012,Ramirez_low_2012}.
In their work they assume that there is some physical mechanism which
would limit the potential from above such that
\(V(\phi)<M_\mathrm{GUT}^4\), and thus find that
\(V(\phi)\ll\dot{\phi}^2\) at early times. Our results therefore place their observations in a more 
generic setting.

\subsection{Comparison with equipartition initial conditions}
\label{sec:comparison}

An alternative method for setting the initial conditions for inflation
models has been proposed by Boyanovsky, de Vega and Sanchez (BVS) in
\citep{boyanovsky_cmb_2006}. Their work also shows that the 
low-multipole suppression of the scalar power spectrum is the result
of a brief period of fast-roll inflation prior to the standard
slow-roll regime. They arrange for such a fast-roll period by assuming
`equipartition initial conditions'. In this approach, the initial
conditions are set at a time \(t=\teq\), when there is approximate
equipartition between the kinetic and potential energy of the
inflaton:
\begin{equation}
  \frac{1}{2}\phideq^2 \sim V(\phieq).
  \label{eqn:bvsbc}
\end{equation}
Almost by definition, the subsequent evolution will generically
exhibit a (brief) period of fast-roll inflation, before entering a
slow-roll phase (which is an attractor solution). BVS verified this 
generic behaviour for a wide range of chaotic and new inflation 
potentials.

\begin{figure}[t!]
  \includegraphics{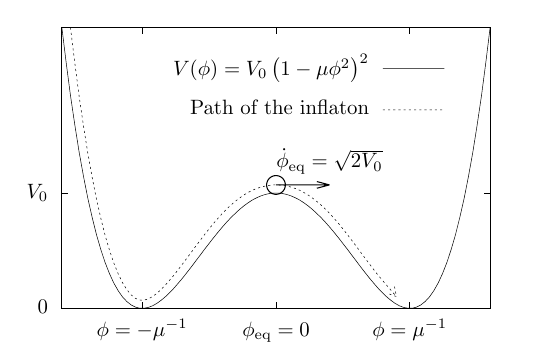}
  \caption{An illustration of the ``equipartition'' initial conditions
      imposed by Boyanovsky, de Vega and Sanchez, for the new inflation 
      potential \(V_0{\left(1-\mu^2\phi^2\right)}^2\). The initial 
      conditions are set at some time \(\teq\) with values denoted by a 
      subscript `eq'. The field is set with \(\phieq=0\) and a velocity 
      \(\phideq= \sqrt{2V_0}\) so that the energy is partitioned equally 
      between kinetic and potential energy: \(V(\phieq) =
      \frac{1}{2}\phideq\). }
      \label{fig:figure_BVS_initial_conditions}
\end{figure}

To address these issues further, we consider in more detail the main
inflationary model used by BVS to illustrate their generic findings.  
The model is spatially-flat, and uses a `new inflation' potential:
\begin{equation}
  V(\phi) =V_0{\left(1-\mu\phi^2\right)}^2.
\end{equation}
The initial conditions are set by choosing \(\phi(\teq)=\phieq=0\) with 
an equipartition of kinetic and potential energies so that
\(\dot{\phi}(\teq)=\phideq=\sqrt{2V_0}\).
\Fref{fig:figure_BVS_initial_conditions} illustrates this
diagrammatically.  The values of \(V_0\) and \(\mu\) are tuned so that the 
power spectrum has an appropriate index \(n_s\) and the correct number 
of \(e\)-folds are generated.

We now interpret this methodology using our formalism. Kinetic initial
conditions will inevitably lead to a time \(\teq\) when the
equipartition condition~\eqref{eqn:bvsbc} is satisfied. Thus, rather
than considering~\eqref{eqn:bvsbc} as a fundamental physical principle, 
it should be regarded as a natural consequence of kinetic initial
conditions. Moreover, this is true independently of any tuning that we
apply to \(\mu\) or \(V_0\) and indeed of the potential.  Moreover,
demanding that~\eqref{eqn:bvsbc} is satisfied at \(\phi=0\) corresponds 
to choosing a specific value of \(\phip\) such that the field 
arrives at \(\phi=0\) with an equipartition of kinetic and potential
energy (This is illustrated in
\Fref{fig:figure_BVS_initial_conditions}).  If, however, \(\phip\) is
required to lie within a certain range of values in order to produce
an appropriate number of \(e\)-folds of inflation, then the choice
\(\phieq=0\) may be inconsistent with kinetic dominance. This can only
be resolved if one is free to choose a general point \(\phieq\) for the 
position of equipartition.  

\section{When is kinetic dominance not the case?}
\label{sec:When_is_kinetic_dominance_not_the_case?}
Having looked in detail at the consequences of kinetic dominance, we
enumerate the instances in which it does not hold. If kinetic
dominance is not the case, then we can conclude that either:
\begin{enumerate}
    \item as one moves backward in time \(\dot{\phi}\to 0\), and the
        inflaton tends to a constant value; or
    \item there is no epoch before which we can say \(\dot{\phi}\ne 0\), 
        and the inflaton continues to oscillate as one moves backward in 
        time.
\end{enumerate}
We shall examine each of these cases in turn.

\subsection{Resting inflaton: \(\dot{\phi}\to 0\)}

\subsubsection{No auxiliary fluids: eternal de Sitter}
We shall begin by considering the case with no auxiliary fluids, for
which the Friedmann~\eqref{eqn:Friedmann_mod} and Klein--Gordon~\eqref{eqn:Klein_Gordon_mod} 
equations take the form:
\begin{align}
  H^2 
  &=
  \frac{1}{3\m^2}\left(\tfrac{1}{2}\dot{\phi}^2 + V(\phi) \right),
  \label{eqn:Friedmann_dotphi_to_zero_no_fluid} 
  \\
  0
  &=
  \ddot{\phi} +3\dot{\phi}H + V^\prime(\phi).
  \label{eqn:Klein_Gordon_dotphi_to_zero_no_fluid}
\end{align}
If \(\dot{\phi}\to 0\) as one moves backward in time, then the inflaton 
\(\phi\) and field \(V(\phi)\) tend to constant values \(\phiz\) and 
\(V(\phiz)\equiv V_0\). By examining the first of the above equations, 
one can see that \(H\) tends to a constant value:
\begin{equation}
  H\to H_0 \equiv \sqrt{\frac{V_0}{3\m^2}}.
\end{equation}
Thus, such a universe exhibits an eternal de Sitter phase (edS) as
\(t\to-\infty\).

By examining the Klein-Gordon
equation~\eqref{eqn:Klein_Gordon_dotphi_to_zero_no_fluid}, one can see that the 
only nonzero term remaining is \(\frac{\d{}}{\d{\phi}}V(\phi)\). From this 
one can conclude that the inflaton must come to rest on an extremum of
the potential. It is straightforward to show that the dynamical
equations then have the following asymptotic solution as \(t \to
-\infty\):
\begin{align}
  \phi(t)
  &=
  \phiz \pm A \exp(\alpha t),\\
  H(t)
  &=
  \sqrt{\frac{V(\phiz)}{3\m^2}} \equiv H_0,\\
  a(t)
  &=
  B e^{H_0t},
\end{align}
where \(A\) and \(B\)  are arbitrary constants, and \(\alpha\) is a real, 
positive solution to the quadratic equation:
\begin{equation}
  \alpha^2 + 3H_0\alpha + \left.
  \frac{{d^2}V}{{d}\phi^2}\right|_{\phi=\phiz}=0.
\end{equation}
This solution is discussed in more detail by 
\citet{destri_preinflationary_2010}. An example of this is ``Hilltop
inflation'' \citep{linde_1982,albrecht_1982}. 

It should be noted that these solutions are not generic, as these
solutions are rolling away from a position of unstable equilibrium:
going backward in time, any small perturbation causes the inflaton to
overshoot the extremum and move on to a kinetically-dominated phase.

\begin{figure}[t!]
  \includegraphics{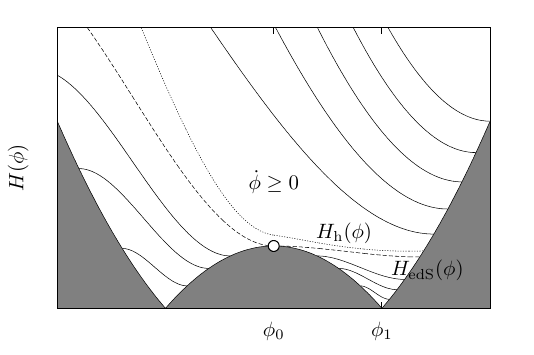}
  \caption{Schematic of the Hubble parameter against \(\phi\) in the
      Hamilton-Jacobi representation. The potential \(V(\phi)\) is the
      same potential as in
      \protect\Fref{fig:figure_BVS_initial_conditions}. The solid curves 
      represent the portions of the solutions to~\protect\eqref{eqn:Hamilton_Jacobi_H} 
      with \(\dot{\phi}\ge0\). The 
      shaded region is defined by \(H<\sqrt{V/3\m^2}\). Solutions meet the 
      shaded region with zero gradient. Solutions within the white 
      region are unique: they do not cross over. The right hand side of
      the dashed curve represents a universe entering at  \(\phi=\phiz\), 
      (\(t=-\infty\)) in an eternal de Sitter phase. The left hand side of
      the dashed curve indicates a universe entering in a kinetically 
      dominated phase before settling into a de Sitter phase at 
      \(t=+\infty\).
  }
  \label{fig:figure_edS}
\end{figure}

One can demonstrate the above statement formally by considering
equations~\eqref{eqn:Klein_Gordon_dotphi_to_zero_no_fluid} and~\eqref{eqn:Friedmann_dotphi_to_zero_no_fluid} 
in the Hamilton--Jacobi 
representation:
\begin{align}
  {\left(\frac{\d{H}}{\d{\phi}}\right)}^2
  &=
  \frac{3H^2}{2\m^2} - \frac{V(\phi)}{2\m^4},
  \\
  \frac{\d{H}}{\d{\phi}}
  &=
  -\frac{\dot{\phi}}{2\m^2}.
\end{align}
The Hamilton--Jacobi representation is valid in the periods in which
\(\phi(t)\) is monotonic; i.e. \(\dot{\phi}\) does not change sign. If one 
assumes that \(\dot{\phi}>0\), then the first of the above equations 
reads:
\begin{equation}
  \frac{\d{H}}{\d{\phi}} 
  = 
  -\sqrt{\frac{3H^2}{2\m^2} - \frac{V(\phi)}{2\m^4}}.
  \label{eqn:Hamilton_Jacobi_H}
\end{equation}
Solutions to this equation are plotted in \Fref{fig:figure_edS}, which
demonstrates the following facts: 
\begin{itemize}
  \item There is a region in which solutions cannot exist, since by
      the Friedmann
      equation~\eqref{eqn:Friedmann_dotphi_to_zero_no_fluid} we find that 
      \(H>\sqrt{V(\phi)/3\m^2}\).
  \item  By equation~\eqref{eqn:Hamilton_Jacobi_H} that solutions meet
      this region with zero gradient.
  \item Outside of this region, the right hand side of equation~\eqref{eqn:Hamilton_Jacobi_H} 
      is Lipschitz continuous \citep[see Appendix~\ref{sec:uniqueness_theorem},
      or][]{agarwal_1993}, hence solutions do not cross over in the
      white region of the graph.
\end{itemize}

Consider the eternal de Sitter solution \(\HedS(\phi)\), plotted 
as the right hand half of the dotted line in \Fref{fig:figure_edS}.
This solution has the property that as \(\phi\to\phiz\),
\(\dot{\phi}\propto\frac{\d{H}}{\d{\phi}}\to0\). Consider also a solution 
\(\Hh\) which at some value \(\phi_1\) is greater than the edS solution, 
\(\Hh(\phi_1) \ge \HedS(\phi_1)\). By uniqueness, this will remain 
greater than \(\HedS\) within the white region of the graph. Both 
\(\HedS\) and \(\Hh\) satisfy equation~\eqref{eqn:Hamilton_Jacobi_H}. 
Taking the difference of these two equations and integrating from 
\(\phi_1\) to \(\phiz\) shows:
\begin{align}
  &\Hh(\phiz)-\HedS(\phiz)= \nonumber\\
  &\Hh(\phi_1)-\HedS(\phi_1) \nonumber\\
  & + \int_{\phiz}^{\phi_1} 
  \sqrt{\frac{3\Hh^2}{2\m^2} - 
  \frac{V(\phi)}{2\m^4}}-\sqrt{\frac{3\HedS^2}{2\m^2} - 
  \frac{V(\phi)}{2\m^4}}\:d\phi \nonumber \\
  &>\Hh(\phi_1)-\HedS(\phi_1)>0.
\end{align}
We thus find \({\d{\Hh}}/{\d{\phi}}\ne0\) at \(\phi=\phiz\), and thus does not 
represent an eternal de Sitter phase. A similar argument holds for 
solutions \(\Hl(\phi)\) which start out less than \(\HedS\), only these 
must collide with the dark region. Upon this collision, \(\dot{\phi}\) 
changes sign, causing the diagram to reflect about the \(\phi=0\) axis.

\Fref{fig:figure_edS} reveals an interesting second eternal de Sitter
solution. The right-hand half of the dashed line indicates the
previously discussed phase: a universe which emerges at \(t=-\infty\) in 
an inflating state with the inflaton slowly rolling off an extremum in 
the potential, exiting the inflationary phase when the inflaton
oscillates about the bottom of the potential.  The left hand half of
the dashed line indicates a universe which emerges at \(t=0\) in a
kinetically dominated phase before the inflaton rests on top of the
extremum, settling into an eternal de Sitter phase as \(t\to\infty\).

\subsubsection{Auxiliary fluids}
The presence of an auxiliary fluid makes it slightly easier to
engineer a solution in which \(\dot\phi\to0\), as one does not require
that the inflaton comes to rest on an extremum of the potential.

In the limit that \(a\to 0\), the \(\rho_i\) with the largest \(w_i\)
dominates over all of the others. The relevant equations are therefore
the Friedmann~\eqref{eqn:Friedmann_mod} and
Klein-Gordon~\eqref{eqn:Klein_Gordon_mod} equations, with a single fluid \(\rho_w\):
\begin{align}
  H^2
  &=
  \frac{1}{3\m^2}\left(\tfrac{1}{2}\dot{\phi}^2
  + V(\phi)
  + \rho_w \right),
  \label{eqn:Friedmann_dotphi_to_zero} \\
  0
  &=
  \ddot{\phi} +3\dot{\phi}H + V^\prime(\phi).
  \label{eqn:Klein_Gordon_dotphi_to_zero}
\end{align}
In the first of these, if \(\phi\to\phiz\), then as \(a\to0\) on the
right-hand side one only need worry about \(\rho_w\). Since
\(H=\frac{\d{}}{\d{t}}\log a\), and \(\rho_w\propto a^{-3(1+w)}\), one
finds:
\begin{align}
  \int\frac{\d{\log a}}{\sqrt{\rho_w}}
  &=
  \int\frac{t}{\sqrt3\m},
  \\
  \Rightarrow \qquad \rho_w
  &=
  \frac{4\m^2}{3{(1+w)}^2} \frac{1}{t^2}.
  \label{eqn:rho_dotphi_to_zero}
\end{align}
We may then use the above equation along
with~\eqref{eqn:Friedmann_dotphi_to_zero} to show that:
\begin{equation}
  H = \frac{2}{3(1+w) t}
  \quad
  \Rightarrow \quad a\propto t^{2/3(1+w)}.
  \label{eqn:H_dotphi_to_zero}
\end{equation}
We can now solve~\eqref{eqn:Klein_Gordon_dotphi_to_zero} for \(\phi(t)\).  
As \(\phi\to\phiz\), we may assume the potential term tends to a 
constant value \(\prm{V_0}\). Selecting the solution which tends to a 
constant, one finds:
\begin{equation}
  \phi(t) = \phiz -\frac{\prm{V_0}(w+1)}{2(w+3)}t^2.
\end{equation}
This constitutes a solution which is present for any potential with
auxiliary fluids. In general one can find a specific solution with
\(\phi\to\phiz\) for any given \(\phiz\). However, for any small
perturbation from this solution, one arrives back at kinetic
dominance. The kinetically dominated solutions are generic.

\subsection{Pathological oscillations: \(\dot{\phi}=0\)}
\label{sec:pathological_oscillations}
We now turn to the case where there is no epoch prior to which \(\phi\) 
is monotonic. The inflaton continues to oscillate endlessly as 
\(a\to0\). 

It is easy to engineer potentials that lead to this behaviour. In the
case with no auxiliary fields, the limiting forms of \(\dot{\phi}(t)\)
and \(\phi(t)\) are given by equations~\eqref{eqn:dotphi_of_t_flat}
and~\eqref{eqn:phi_of_t_flat}. We may solve these to find the relationship 
between \(\dot{\phi}\) and \(\phi\):
\begin{equation}
  \dot{\phi}^2 = \exp\left(\frac{\sqrt{6}}{\m} |\phi-\phip|\right).
\end{equation}
If one chooses a potential that grows faster than the right hand side
of this equation, then the universe cannot be kinetically dominated.
We are thus forced to the conclusion that in such a universe either
\(\dot{\phi}\to 0\) or there is no epoch before which \(\dot{\phi}\ne 0\).  
Typically, if one examines the numerical solutions of such equations, 
one sees that the inflaton oscillates at a faster and faster rate, 
with greater and greater amplitude until the numerical limit of the 
solver is reached.

These solutions are therefore somewhat pathological, though for the
cases where they occur, kinetic dominance is not the generic solution.

\section{Conclusions}
\label{sec:Conclusions}

We have shown that, if quantum gravitational effects are ignored, the
coupled evolution equations for the inflaton field \(\phi(t)\) and the
Hubble parameter \(H(t)\) in generic homogeneous and isotropic
single-field inflation models imply that a universe beginning with a
steadily moving inflaton (\(|\dot{\phi}| > \vellim > 0\) as \(a\to 0\), 
for some positive constant \(\vellim\)) generically emerges from an 
initial singularity in a non-inflating, kinetically dominated state
(\(\dot{\phi}^2 \gg V(\phi)\)).  In this kinetic-dominated regime, one
obtains simple analytical solutions for \(\phi(t)\) and \(H(t)\), which 
are independent of the form of the inflaton potential \(V(\phi)\) and of 
the presence of auxiliary fluids such as matter, radiation, dark 
energy or spatial curvature. These solutions provide a simple means of 
setting the initial conditions for such inflation models, from which
numerical integration of the evolution equations may proceed.

For illustration, we applied this `kinetic' procedure for setting
initial conditions to spatially-flat polynomial and exponential
inflation models.  By making an appropriate choice of the time \(\ti\)
at which the initial conditions are set, and the single free parameter
\(\phip\) in the analytic kinetic-dominated solution, all models produce 
an amount of inflation compatible with observations.  The background
evolution in each case displays a generic behaviour.  Following a
non-inflating period of kinetic dominance, \(H(t)\), \(\phi(t)\) and their 
time-derivatives continue to decrease until one obtains approximate
equipartition \(\dot{\phi}^2 \sim V(\phi)\) between the kinetic and
potential energies of the inflaton. This marks the onset of a
(typically brief) period of fast-roll inflation, which turns into a
slow-roll [\(\dot{\phi}^2 \ll V(\phi)\)] or power-law inflation phase.
At the end of the slow-roll phase, the inflation quickly moves towards
a minimum of the potential, about which it executes a decaying
oscillation.

We calculated the approximate spectrum of scalar perturbations 
for the polynomial and exponential models and find, in both cases,
that it contains less power at low- and high-\(k\) values than would be 
expected from a power-law behaviour. The low-\(k\) effect is a
generic consequence of the kinetic initial conditions, for any
consistent inflaton potential or spatial curvature, resulting in
particular from the brief period of fast-roll inflation that they
imply.  The damping of power on large scales may provide an
explanation for the low-\(\ell\) falloff in the matter and CMB power
spectra seen in recent cosmological observations.

We also compared our kinetic initial conditions with an alternative
proposal by Boyanovsky, de Vega and Sanchez
\citep{boyanovsky_cmb_2006} that inflationary initial conditions
should be set by assuming approximate equipartition between the
kinetic and potential energy of the inflaton.  In the context of
kinetic initial conditions, approximate equipartition is not a
fundamental physical principle, but merely an inevitable consequence.
Moreover, by considering a particular model used by BVS, we 
demonstrate that assigning equipartition initial conditions with an 
arbitrary initial value for the inflaton field can lead to 
inconsistency with kinetic initial conditions.

Finally we enumerated the universes which do not have a steadily
moving inflaton, and have shown that these are special cases, distinct
from the generic kinetically dominated case.

\section*{Acknowledgements}
We thank Norma Sanchez for providing useful comments on a very early version of
this paper originally drafted in June 2009. We also thank the two anonymous
referees for numerous useful comments, and Anthony Challinor for discovering an
error in the proof (now corrected) during W~H thesis assesment. S~D~B
thanks the Isaac Newton Trust and the Sunburst Fund for their support. W~H
thanks STFC for their support.

\appendix

\section{Uniqueness theorem}
\label{sec:uniqueness_theorem}
We shall now prove that the solutions to the initial value problem of
the master equation~\eqref{eqn:master_eq}:
\begin{align}
  \frac{\d{y}}{\d{\psi}}
  &=
  \sqrt{1-e^{-2y}} - \frac{\d{}}{\d{\psi}}\log \sqrt V,
  \label{eqn:IVP1}
  \\
  y(\psiz)
  &=
  y_0>0
  \label{eqn:IVP2},
\end{align}
are unique within any finite interval \(\psi\in[\psiz,\psi_1]\); i.e, if 
two positive solutions intersect at a point, then they intersect
everywhere.

We begin by putting a lower bound on \(y\) in the interval
\([\psiz,\psi_1]\): From assumption~\eqref{eqn:conditions}, we know
\(\dot{\phi}^2>\vellim^2>0\). If we unpack the definition of \(y\) using
equations~\eqref{eqn:y_def},~\eqref{eqn:Ntrans},~\eqref{eqn:tau_def}
and~\eqref{eqn:Friedmann_c}, one finds:
\begin{equation}
  y 
  = 
  \frac{1}{2}\log
  \left(\frac{\frac{1}{2}\dot{\phi}^2 + V(\phi)}{V(\phi)}\right) 
  > 
  \frac{\vellim^2}{4\Vm},
\end{equation}
where \(\Vm\) is the maximal value of \(V(\psi)\) in the interval
\([\psiz,\psi_1]\). With this in hand we may prove the uniqueness of
solutions of the initial value problem~\eqref{eqn:IVP1},~\eqref{eqn:IVP2} 
using standard techniques.  For a good reference of such techniques the 
reader should consult the text by Agarwal et al.\
\citep{agarwal_1993}.  In this case, we shall prove it using {\em Peano 
iteration}.

If one assumes that \(y(\psi)\) and \(z(\psi)\) are two distinct
solutions, then their difference satisfies:
\begin{equation}
  \frac{\d{}}{\d{\psi}}(y-z)
  =
  \sqrt{1-e^{-2y}} - \sqrt{1-e^{-2z}}.
  \label{eqn:master_diff}
\end{equation}
If in addition one assumes they meet at a common point \(\psi_0\), so
that \(y(\psi_0)=z(\psi_0)\), then integrating away from this position
yields:
\begin{align}
  \abs{y(\psi)-z(\psi)}
  =&
  \abs{\int_{\psi_0}^\psi \sqrt{1-e^{-2y}}
  - \sqrt{1-e^{-2z}}\:\:d\psi}
  \nonumber\\
  &
  \le \int_{\psi_0}^\psi \abs{\sqrt{1-e^{-2y}}
  - \sqrt{1-e^{-2z}}}d\psi.
  \label{eqn:ineq_1}
\end{align}
A generic property of the function \(f(y)=\sqrt{1-e^{-2y}}\) is that
in the interval \(\left[\phantom(\frac{\vellim^2}{4\Vm},\infty\right)\phantom]\) it is {\em Lipschitz 
continuous\/}:
\begin{equation}
  \abs{\sqrt{1-e^{-2y}} - \sqrt{1-e^{-2z}}} \le L\abs{y-z},
  \label{eqn:Lipschitz}
\end{equation}
where \(L\) is the {\em Lipschitz constant}, taking the value:
\begin{equation}
  L
  = 
  \frac{\exp{\left(-\frac{\vellim^2}{2\Vm}\right)}}
  {\sqrt{1-\exp{\left(-\frac{\vellim^2}{2\Vm}\right)}}} 
  > 0.
\end{equation}
Applying Lipschitz continuity~\eqref{eqn:Lipschitz} to the inequality
in~\eqref{eqn:ineq_1} gives:
\begin{equation}
  \abs{y(\psi)-z(\psi)} 
  \le
  L\int_{\psi_0}^\psi\abs{y(\psi)-z(\psi)}\d{\psi}.
  \label{eqn:ineq_2}
\end{equation}
Further, if the maximum value of the difference of \(|y-z|\) between
\(\psi_0\) and \(\psi\) is \(\Delta\), then the above implies:
\begin{equation}
  \abs{y(\psi)-z(\psi)} 
  \le
  L\Delta\abs{\int_{\psi_0}^\psi \d{\psi}}= 
  L\Delta\abs{\psi-\psi_0}.
\end{equation}
Applying this inequality back into~\eqref{eqn:ineq_2} shows:
\begin{equation}
  \abs{y(\psi)-z(\psi)} 
  \le
  L^2\Delta\int_{\psi_0}^\psi\abs{\psi-\psi_0}\d{\psi} =
  L^2\Delta\frac{\abs{\psi-\psi_0}^2}{2!}.
\end{equation}
Applying this back into~\eqref{eqn:ineq_2} yields:
\begin{equation}
  \abs{y(\psi)-z(\psi)} 
  \le
  L^3\Delta\frac{\abs{\psi-\psi_0}^3}{3!},
\end{equation}
and by induction on \(n\in\mathbb{N}\) we find that:
\begin{equation}
  \abs{y(\psi)-z(\psi)} 
  \le
  L^n\Delta\frac{\abs{\psi-\psi_0}^n}{n!}.
\end{equation}
As \(n\to\infty\) the term on the right-hand side drops to \(0\), and
therefore \(|y(\psi)-z(\psi)|=0\). Thus, if \(y\) and \(z\) are equal at 
some point \(\psi_0\), then they are equal at all points \(\psi\) within 
any finite interval \([\psiz,\psi_1]\). Two separate solutions cannot 
`cross over', and if one positive solution \(f\) is initially less than 
a second solution \(h\) at \(\psiz\), \(f(\psiz)<h(\psiz)\), then 
\(f(\psi_1)<h(\psi_1)\) for any finite \(\psi_1\).

\bibliography{kineticdominance}

\end{document}